\documentclass[11pt,letterpaper]{article}

\def\makinhome{/home/makin}
\makeatletter
\begingroup\endlinechar=-1\relax
    \everyeof{\noexpand}%
    \edef\x{\endgroup\def\noexpand\homepath{%
        \@@input|"kpsewhich --var-value=HOME" }}\x
\makeatother

\ifx\homepath\makinhome
    \newcommand{\stydir}{../../stys}
    
    \newcommand{\tikzdir}{../../tikzpics/IPDiMC}
    \newcommand{\figdir}{../../../jpgs/IPDiMC}
\else
    \newcommand{\stydir}{stys}
    
    \newcommand{\tikzdir}{tikzpics/IPDiMC}
    \newcommand{\figdir}{jpgs/IPDiMC}
\fi

\providecommand{\stydir}{../stys}

\providecommand{\tikzdir}{../tikzpics}

\usepackage{%
	amsmath,amsfonts,
	amssymb,amsthm,graphicx,rotating,booktabs,
	versions, ifthen,siunitx,
}
\usepackage[table]{xcolor}
\usepackage{xcolor}
\usepackage{pgf,tikz,pgfplots} 
\usetikzlibrary{%
 	positioning,calc,patterns,shapes,arrows,intersections,backgrounds,fit,shadings,decorations.text,shadows,fadings,bayesnet,
}
\usepgfplotslibrary{colorbrewer,colormaps,fillbetween,groupplots}

\ifdefined\rvmacroize
\else
	\makeatletter
	\input{\stydir/jgmstd.sty}
	\makeatother
\fi

\newboolean{INCLFIGS} 
\setboolean{INCLFIGS}{true}

\usepackage{caption,subfig}
\newcommand{\captioning}[2]{\caption{{\bf #1} {#2}}}
\ifthenelse{\boolean{INCLFIGS}}{
	\captionsetup[subfloat]{font={Large,sf},labelformat=simple,labelsep=none}
}{
	\captionsetup[subfloat]{labelformat=empty,labelsep=none}
}

\newcommand{\colorprovide}[2]{\@ifundefinedcolor{#1}{\colorlet{#1}{#2}}{}}

\makeatletter
\input{\stydir/IMSL.sty}
\makeatother

\usepackage{multirow,authblk,url} 

\pgfplotsset{compat=1.17}

\input{IPDiMC.sty}

\newcommand{\TableDataSets}{
    \begin{table}[ht]
        \begin{minipage}{\textwidth}
            \centering\small
            \begin{tabular}{@{}lllcrrrr@{}}
                name
                & area
                & task
                & $N_\text{in}$
                & $N_\text{out}$
                & $M_\text{in}$
                & $M_\text{fw}$
                & $N_\text{pre-movement}$\\
                \toprule
                \mcmaze
                & M1, PMd
                & center-out reaches %
                & 137
                & 45
                & 140
                & 40
                & $50$\\
                \cite{Churchland2010b,Churchland2012} && w/obstacles &&&&&\\
                \midrule    
                \mcrtt
                & M1
                & point-to-point reaches
                & 98
                & 32
                & 120
                & 40
                & arbitrary\\
                \cite{ZenodoNoLink2017,Makin2018} && &&&&&\\
                \midrule
                \areabump
                & S2
                & reach to visual target
                & 49
                & 16
                & 120
                & 40
                & $20$\\
                \cite{Chowdhury2020} && w/displacement &&&&&\\
                \midrule    
                \dmfcrsg
                & DMFC
                & ready-set-go time-
                & 40
                & 14
                & 300
                & 40
                & $300\pm30$\\
                \cite{Sohn2019}  && interval matching &&&&&\\
                \bottomrule
            \end{tabular}
            \captioning{Description of data sets.}{%
            M1 = primary motor area; PMd = premotor dorsal; S2 = secondary somatosensory cortex; DMFC = dorsolateral medial frontal cortex (including supplementary eye field, supplementary motor area, and pre-SMA); $N_\text{in}$ = number of observed (``held-in'') neurons; $N_\text{out}$ = number of unobserved (``held-out'') neurons; $M_\text{in}$ = number of observed (``held-in'') samples; $M_\text{fw}$ = number of future (``forward'') unobserved samples; $N_\text{pre-movement}$ = number of samples of the trial before movement onset (for \mcrtt, trials were not aligned to movements).
            For all data sets, spike bins are 5 ms.%
            }
            \label{tbl:databreakdown}
        \end{minipage}
    \end{table}
}

\newcommand{\FigTraining}{
    \begin{figure}[!t]
     \centering
        \TikzTraining
        \captioning{Training Methodology.}{
        A single trial's data correspond to an array of size (number of neurons [$N$]) $\times$ (number of samples [$M$]) (top left).
        To construct input data (right), the ``held-in'' block of spike counts is extracted, zeroed at random locations (``masked''; light blue), and zero-padded out to the total trial length $M$ (upper right).
        This array of integers is then passed through the model, which predicts (real-valued) firing rates for all held-out and held-in neurons, for the entire length of the trial (right).
        To construct output targets (left), the mask is ``inverted'' and applied to the original data array (lower left), meaning that only the ``missing'' portions of the input, including the held-out neurons, are retained.
        The inverted mask is also applied to the output of the model (lower right), and the loss is computed on the retained elements. 
        }
        \label{fig:TrainingMethodology}
    \end{figure}
}

\newcommand{\FigArchitectures}{
    \begin{figure}[!t]
        \subfloat[][]{%
            \label{subfig:RNN}
            \begin{minipage}[t]{0.30\linewidth}\vspace{0in}
                \TikzRNNF
            \end{minipage}
        }%
        \hspace{0.015\linewidth}
        \subfloat[][]{%
            \label{subfig:transformer}
            \begin{minipage}[t]{0.3\linewidth}\vspace{0in}
                \TikzTransF
            \end{minipage}
          \label{fig:sfig2}
        }%
        \hspace{0.015\linewidth}
        \subfloat[][]{
            \label{subfig:TERN}
            \begin{minipage}[t]{0.30\linewidth}\vspace{0in}
                \TikzTERN
            \end{minipage}
        }
        \captioning{Model Architectures.}{\subcapref{RNN} \RNN\ is a multi-layer, bidirectional, recurrent neural network with gated recurrent units (GRUs), followed by a single non-recurrent (feedforward) layer.
        \subcapref{transformer} \transformer\ is the encoder half of a Transformer with multiple self-attention layers and a final feedforward layer.
        This model is similar to the Neural Data Transformer (NDT) \cite{Ye2021}.
        \subcapref{TERN} \TERN\ (Transformer Encoder over Recurrent Network) is a multi-layer, bidirectional RNN followed by a single layer of a transformer encoder, again terminating in a (static) feedforward layer.
        Input and output data for all architectures are the same.
         }
        \label{fig:modelArchitectures}
    \end{figure}   
}

\newcommand{\TableVelocityPrediction}{
    \begin{table}[ht]
    \begin{minipage}{\textwidth}
    \begin{center}
    \small
    \begin{tabular}{@{}llllcc@{}}
    \multirow{3}{*}{Dataset} & \multicolumn{3}{c}{Architecture}   & \\
                            \cmidrule(lr){2-4} 
                             & \multicolumn{1}{l}{\RNN}           & \multicolumn{1}{l}{\transformer}      & \multicolumn{1}{l}{\TERN} & smoothed & raw \\
                            & & & & spike counts & spike counts\\
    \midrule
    \mcmaze                 & \multicolumn{1}{l}{$0.8886\pm0.0062$} & \multicolumn{1}{l}{$0.8740\pm0.0061$} & \multicolumn{1}{l}{\boldsymbol{$0.9003\pm0.0025$}} & 0.6238 & 0.0949\\
    \mcrtt                  & \multicolumn{1}{l}{$0.5588\pm0.0083$} & \multicolumn{1}{l}{$0.4575\pm0.0222$} & \multicolumn{1}{l}{\boldsymbol{$0.5753\pm0.0094$}} & 0.4142 &0.0432\\
    \areabump              & \multicolumn{1}{l}{$0.8561\pm0.0122$} & \multicolumn{1}{l}{$0.8152\pm0.0163$} & \multicolumn{1}{l}{\boldsymbol{$0.8612\pm0.003$}}  & 0.5736 & 0.1313\\
    \bottomrule
    \end{tabular}
    \captioning{Hand-velocity $R^2$ on the test partition.}{%
    The mean coefficient of determination ($\pm$ standard deviation) for velocity predictions is given for each data set, across the best ten instances of each architecture.
    Predictions are made with an affine transformation of the firing rates of both held-in and held-out neurons that have been inferred by the models.
    The final column quantifies predictions made with an affine transformation of the \emph{inputs}, rather than outputs, of the models; i.e., the spike counts rather than the inferred firing rates.
    }
        \label{tbl:velocityPrediction}
    \end{center}
    \end{minipage}
    \end{table}
}

\newcommand{\TableLeaderboard}{   
    \begin{table}[!htb]
        \centering\small%
        \begin{tabular}{@{}lllll@{}}
            & {\fontfamily{qcr}\selectfont \mcmaze} & {\fontfamily{qcr}\selectfont \mcrtt} & {\fontfamily{qcr}\selectfont \areabump} & {\fontfamily{qcr}\selectfont \dmfcrsg} \\
            \toprule
            smoothing & 0.211 & 0.147 & 0.154 & 0.12 \\ 
            GPFA & $0.187\pm0.001$ & $0.155\pm0.000$ & $0.168\pm0.000$ & $0.118\pm0.000$ \\ 
            SLDS & $0.219\pm0.006$ & $0.165\pm0.004$ & $0.187\pm0.005$ & $0.120\pm0.001$ \\ 
            NDT & $0.329\pm0.005$ & $0.160\pm0.009$ & $0.267\pm0.003$ & $0.162\pm0.009$ \\ 
            AutoLFADS & $0.346\pm0.005$ & $0.192\pm0.003$ & $0.259\pm0.001$ & $\mathbf{0.181\pm0.001}$ \\ 
            \transformer$^*$ & $0.360\pm0.003$ & $0.071\pm0.126$ & $0.261\pm0.006$ & $0.150\pm0.059$ \\ 
            \RNN$^*$ & $0.364\pm0.003$ & $0.194\pm0.019$ & $0.281\pm0.003$ & $0.180\pm0.003$ \\ 
            \textbf{\TERN$^*$} & $\mathbf{0.373\pm0.002}$ & $\mathbf{0.207\pm0.002}$ & $\mathbf{0.288\pm0.004}$ & $0.174\pm0.024$ \\
            \bottomrule
        \end{tabular}
    \captioning{State of the art in co-smoothing.}{Comparison with the top models submitted to the Neural Latents Benchmark '21 \cite{Pei2021}.
    The models introduced in this work are indicated by $^*$.}
    \label{tbl:leaderboard}
    \end{table}
}

\newcommand{\FigDynamicMaskingEffects}{
    \begin{figure}[!htb]
        \centering
        \subfloat[][]{%
            \label{subfig:dynamic_masking_effect_co-bps}
            \begin{minipage}[c]{0.5\textwidth}
                \centering
                \includegraphics[width=0.95\linewidth]{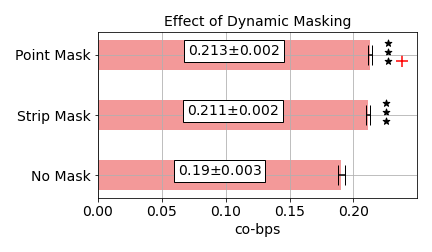}
            \end{minipage}
        }
        \subfloat[][]{%
           \label{subfig:dynamic_masking_effect_velR2}
            \begin{minipage}[c]{0.5\textwidth}
                \centering
                \includegraphics[width=0.95\linewidth]{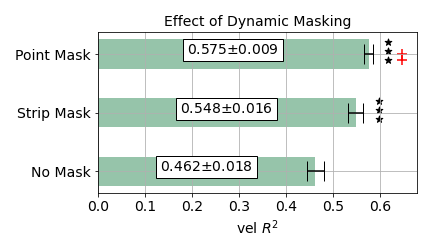}
            \end{minipage}
        }
        \captioning{Effect of dynamic masking on \TERN\ (\mcrtt).}{%
        For each of the three masking options (bars), hyperparameters for the \TERN\ model were optimized from scratch (see \sctname{methods}), and the top 10 (out of 125) models selected.
        Performance (mean with standard deviation) is quantified in terms of \subcapref{dynamic_masking_effect_co-bps} co-bps and \subcapref{dynamic_masking_effect_velR2} velocity predictions (coefficient of determination).
        The ``point mask'' zeros out random samples, whereas the ``strip mask'' zeros out random ``strips'' of three consecutive samples (both for random neurons).
        Point and strip masking were compared with no masking (***$: p<0.0005$, one-sided Wilcoxon rank-sum test, Holm-Bonferroni corrected) and with each other (+$: p<0.05$, ++$: p<0.005$) (two-sided Wilcoxon signed-rank test, Holm-Bonferroni corrected).
        }
        \label{fig:dynamicMaskingEffects}
    \end{figure}
}

\newcommand{\FigOptuna}[2]{
    \begin{figure}[!h]
        \scriptsize%
        \newcommand{\figwidth}{0.38\linewidth}%
        \newcommand{\figheight}{1.5in}%
        \includegraphics[width=\linewidth]{\figdir/mc_rtt_Optuna_#1.png}
        \captioning{Results of Optuna hyperparameter runs (#2).} 
        {The set of parameters for each run are connected by a line, with color indicating the validation co-bps (the optimization criterion).
        Key:\ size of the RNN layers (hidden\_dim), number of RNN layers (n\_layers), the size of the transformer attention and output layers (hidden\_size), the number of transformer layers (n\_layers\_trans), and the size of the transformer feedforward layer (intermediate\_units).
        }
        \label{Optuna_#1}
    \end{figure}
}

\newcommand{\TableHyperparameters}{
    \begin{table}[!htb]
    \centering\small%
    \begin{tabular}{@{}llccc@{}}
    & \textbf{Search range} & \multicolumn{3}{c}{\textbf{Best}}\\
    \cmidrule{3-5}
    & & \RNN & \transformer & \TERN \\
    \midrule
    layers (RNN)            & int (1,6*)    & 3/2/2/3 & - & 2/2/2/1\\
    hidden units (RNN)      & int (32, 256) & 233/176/195/170   & -  & 133/98/252/198\\
    layers (trans)          & int (1,6*)    & -     & 3/6/6/5  & 1/1/1/1\\
    hidden units (trans)    & int (32, 512) & -     & 112/79/32/58  & 142/95/294/79\\
    feedfwd units (trans) & int (32, 512)   & -     & 44/460/477/252   & 449/431/164/509\\
    \bottomrule
    \end{tabular}
    \caption{Hyperparameter tuning on \mcrtt/\mcmaze/\areabump/\dmfcrsg.
    The final three columns show the hyperparameter values for the top four models (of each architecture tyoe).
    *For \TERN, the number of RNN layers is capped at 3, and the number of transformer-encoder layers is capped at 1.}
    \label{tbl:hyperparameters}
    \end{table}
}

\tikzset{%
    data/.style={rectangle, draw=Set2-F!80!black, fill=Set2-F!35!white, align=center},
    intersection/.style={coordinate},
}
\tikzset{nowdata/.style={data, minimum width=1.25in}}
\tikzset{indata/.style={data, minimum height=0.5in}}
\tikzset{fwddata/.style={data, minimum width=0.4in}}
\tikzset{outdata/.style={data, minimum height=0.15in}}
\tikzset{%
    block/.style={rectangle, rounded corners, text width=1.0in, draw=black, align=center, draw=Set2-D, fill=Set2-D!50!white},
    intersection/.style={coordinate},
}
\tikzset{%
    rnn_block/.style={block, minimum height=0.4in, draw=Set2-D, fill=Set2-D!70!white},
}
\tikzset{%
    mask/.style={draw=Set2-A!80!black, fill=Set2-A!70!white},
}
\tikzset{%
    trans_block/.style={block, minimum height=0.4in, draw=Set2-B, fill=Set2-B!70!white},
}
\tikzset{%
    ff_block/.style={block, minimum height=0.2 in, draw=Set2-C, fill=Set2-C!70!white},
}
\tikzset{%
    spike/.style={rectangle, minimum height=0.05in, minimum width=0.03in, draw=none, fill=gray!90!white, inner sep=0pt, anchor=north west},
    intersection/.style={coordinate},
}
\tikzset{%
    clipspike/.style={rectangle, clip, minimum height=0.05in, minimum width=0.03in, inner sep=0pt, anchor=north west},
    intersection/.style={coordinate},
}

\tikzset{
    pics/NLBdata/.style n args={4}{code={%
        \begin{scope}[every node/.append style={anchor=center}]

        \node[indata,nowdata,anchor=north west] (dataIN) {};
        \coordinate(-northwest) at (dataIN.north west);
        \node[indata,fwddata, right=0.0in of dataIN] (dataINFWD){};
        \ifthenelse{\equal{#1}{INCLUDEOUT}}{%
            \node[outdata,nowdata,below=0.0in of dataIN] (dataOUT){};
            \node[outdata,fwddata,right=0.0in of dataOUT] (dataOUTFWD){};
            \def\maxchannel{12}
        }{
            \begin{scope}
            \clip (dataIN.south west) rectangle (dataINFWD.north east) {};
            \def\maxchannel{9}
        }

        \ifthenelse{\equal{#3}{spikes}}{%

            \pgfmathsetseed{5}
            \foreach \y in {0,1,...,\maxchannel}{%
                \foreach \x in {1,2,...,5}{%
                    \pgfmathparse{(rand+1)/2*(1.65)}
                        \node[spike, anchor=north west]
                        at ($(dataIN.north west) + (\pgfmathresult in,-0.05*\y in)  $)
                        {};
                }
            }    

        }{
            \ifthenelse{\equal{#3}{firingrates}}{
                \node[anchor=north west] at ($(dataIN.north west) + (-0.05in, 0)$){%
                    \newcommand{\figwidth}{2.37in}%
                    \newcommand{\figheight}{1.28in}%
                    \provideboolean{CLEANXAXIS}\setboolean{CLEANXAXIS}{true}%
                    \provideboolean{CLEANYAXIS}\setboolean{CLEANYAXIS}{true}%
                    \input{\tikzdir/fake_rates}%
                };
            }{
            }

        }  

        \ifthenelse{\equal{#4}{pointmask}}{
            \pgfmathsetseed{4}

            \def\maxchannel{9}
            \foreach \y in {0,1,...,\maxchannel}{%
                \foreach \x in {1,2,...,11}{%
                    \pgfmathparse{(rand+1)/2*(1.25)}
                        \node[spike, mask, draw=none, anchor=north west]
                        at ($(dataIN.north west) + (\pgfmathresult in,-0.05*\y in)  $)
                        {};
                }
            }
        }{
            \ifthenelse{\equal{#4}{inversemask}}{
                \def\maxchannel{9}
                \pgfmathsetseed{4}
                \draw[mask, draw=none, even odd rule]
                    (0,0) rectangle (1.25in, -0.5in)
                    \foreach \y [evaluate={\yy=(-0.05*\y)};] in {0,1,...,\maxchannel}{%
                        \foreach \x [evaluate={\xx=(rand+1)/2*(1.25);}] in {1,2,...,11}{%
                            (\xx in, \yy in) rectangle ++(0.03in, -0.05in)
                        }
                    }
                ;
            }{
            }
        }

        \node (datalabel) at (dataIN) {\tikzpictext};
        \node[indata, fwddata, right=0.0in of dataIN, draw=none, fill=none, #2](masked){#2};
        \coordinate (datatop) at ($(dataIN.north west) + (-0.1in, 0in)$) {};
        \coordinate (datamid) at ($(dataIN.south west) + (-0.1in, 0in)$) {};
        
        \ifthenelse{\equal{#1}{INCLUDEOUT}}{
            \coordinate (databottom) at ($(dataOUT.south west) + (-0.1in, 0in)$) {};
            \coordinate (dataleft) at ($(dataOUT.south west) + (0in, -0.1in)$) {};
            \coordinate (dataright) at ($(dataOUT.south east) + (0in, -0.1in)$) {};
            \coordinate (dataend) at ($(dataOUTFWD.south east) + (0in, -0.1in)$) {};
            \draw[<->] (datamid.center) -- (databottom.center) node[midway, left] {$N_\text{out}$};
            
            \node[outdata, nowdata, below=0.0in of dataIN, draw=none, fill=none, #2] (maskOUT){};
            \node[outdata, fwddata, right=0.0in of dataOUT, draw=none, fill=none, #2] (maskOUTFWD){};
        }{
            \end{scope}
            \coordinate (databottom) at ($(dataIN.south west) + (-0.1in, 0in)$) {};
            \coordinate (dataleft) at ($(dataIN.south west) + (0in, -0.1in)$) {};
            \coordinate (dataright) at ($(dataIN.south east) + (0in, -0.1in)$) {};
            \coordinate (dataend) at ($(dataINFWD.south east) + (0in, -0.1in)$) {};    
        }

        \draw[<->] (datatop.center) -- (datamid.center) node[midway, left] {$N_\text{in}$};
        \draw[<->] (dataleft.center) -- (dataright.center) node[midway, below] {$M_\text{in}$};
        \draw[<->] (dataright.center) -- (dataend.center) node[midway, below] {$M_\text{out}$};

        \path(dataIN.north west) -- (dataINFWD.north east) coordinate[midway](-topmiddle){};
        \ifthenelse{\equal{#1}{INCLUDEOUT}}{%
            \path(dataOUT.south west) -- (dataOUTFWD.south east) coordinate[midway](-bottommiddle){};
            \path(dataIN.north west) -- (dataOUT.south west) coordinate[midway](-middleleft){};
            \path(dataINFWD.north east) -- (dataOUTFWD.south east) coordinate[midway](-middleright){};
        }{
            \path(dataIN.south west) -- (dataINFWD.south east) coordinate[midway](-bottommiddle){};
            \path(dataIN.north west) -- (dataIN.south west) coordinate[midway](-middleleft){};
            \path(dataINFWD.north east) -- (dataINFWD.south east) coordinate[midway](-middleright){};
        }

        \end{scope}
    }}
}

\newcommand{\TikzRNNF}{%
    \usetikzlibrary{quotes}
    \begin{tikzpicture}[node distance=1cm, auto]%
        \footnotesize
        \tikzset{>={Latex[width=1.5mm,length=1.5mm]}}

        \draw pic["spikes"] (INPUT) {NLBdata={}{mask}{spikes}{}};
   
        \node[rnn_block, below=0.35in of INPUT-bottommiddle] (rnnblock) {multi-layer RNN};
        \node[ff_block, below=0.25in of rnnblock] (ffblock) {feed-forward};

        \node (temp) at ($(INPUT-northwest |- ffblock.south) + (0,-0.25in)$) {};
        \draw pic (OUTPUT) at (temp)  {NLBdata={INCLUDEOUT}{}{firingrates}{}};
        \node[below=0.25in of OUTPUT-bottommiddle] (foo) {firing rates};

        \draw[->,ultra thick] (INPUT-bottommiddle) -- (rnnblock);
        \draw[->,ultra thick] (rnnblock) -- (ffblock);
        \draw[->,ultra thick] (ffblock) -- (OUTPUT-topmiddle);
    \end{tikzpicture}%
}

\newcommand{\TikzTransF}{%
    \usetikzlibrary{quotes}
    \begin{tikzpicture}[node distance=1cm, auto]
        \footnotesize
        \tikzset{>={Latex[width=1.5mm,length=1.5mm]}}
        
        \draw pic["spikes"] (INPUT) {NLBdata={}{mask}{spikes}{}};
 
        \node[trans_block, below=0.35in of INPUT-bottommiddle] (transblock) {transformer encoder};
        \node[ff_block, below=0.25in of transblock] (ffblock) {feed-forward};

        \node (temp) at ($(INPUT-northwest |- ffblock.south) + (0,-0.25in)$) {};
        \draw pic (OUTPUT) at (temp)  {NLBdata={INCLUDEOUT}{}{firingrates}{}};
        \node[below=0.25in of OUTPUT-bottommiddle] (foo) {firing rates};

        \draw[->,ultra thick] (INPUT-bottommiddle) -- (transblock);
        \draw[->,ultra thick] (transblock) -- (ffblock);
        \draw[->,ultra thick] (ffblock) -- (OUTPUT-topmiddle);
    \end{tikzpicture}%
}

\newcommand{\TikzTERN}{%
    \usetikzlibrary{quotes}
    \begin{tikzpicture}[node distance=1cm, auto]
        \footnotesize
        \tikzset{>={Latex[width=1.5mm,length=1.5mm]}}

        \draw pic["spikes"] (INPUT) {NLBdata={}{mask}{spikes}{}};

        \node[rnn_block, below=0.35in of INPUT-bottommiddle] (rnnblock) {multi-layer RNN};
        \node[ff_block, below=0.25in of rnnblock] (ffblocka) {feed-forward};

        \node[trans_block, below=0.25in of ffblocka] (transblock) {transformer encoder (1 layer)};
        \node[ff_block, below=0.25in of transblock] (ffblockb) {feed-forward};

        \node (temp) at ($(INPUT-northwest |- ffblockb.south) + (0,-0.25in)$) {};
        \draw pic (OUTPUT) at (temp)  {NLBdata={INCLUDEOUT}{}{firingrates}{}};
        \node[below=0.25in of OUTPUT-bottommiddle] (foo) {firing rates};

        \draw[->,ultra thick] (INPUT-bottommiddle) -- (rnnblock);
        \draw[->,ultra thick] (rnnblock) -- (ffblocka);
        \draw[->,ultra thick] (ffblocka) -- (transblock);
        \draw[->,ultra thick] (transblock) -- (ffblockb);
        \draw[->,ultra thick] (ffblockb) -- (OUTPUT-topmiddle);

    \end{tikzpicture}%
}

\newcommand{\TikzTraining}{%
\begin{tikzpicture}[node distance=1cm, auto]%
    \footnotesize
    \tikzset{>={Latex[width=1.5mm,length=1.5mm]}}

    \draw pic (trainingdata) {NLBdata={INCLUDEOUT}{}{spikes}{}};
    \node[above=0.0in of trainingdata-topmiddle] (trainingdatalabel) {training data};
    \draw pic[below right=1in and 1.5in of trainingdata-topmiddle] (preinput) {NLBdata={}{}{spikes}{}};
    \node[block] (split) 
    at (preinput-middleleft -| trainingdata-bottommiddle)
    {input/target split};

    \draw pic[below left=0.5in and 0.25in of split] (inputmask) {NLBdata={}{mask}{}{pointmask}};
    \node[above=0.0in of inputmask-topmiddle] (trainingdatalabel) {input mask};
    \node[block, mask, below=1.0in of inputmask-bottommiddle] (dynamicmask) {dynamic mask};
    \draw pic (outputmask)
    at ($(inputmask-northwest |- dynamicmask.south) + (0,-1.0in)$)
    {NLBdata={INCLUDEOUT}{mask}{}{pointmask}};
    \node[above=0.0in of outputmask-topmiddle] (trainingdatalabel) {output mask};

    \node[block, mask]
    at (outputmask-middleright -| preinput-bottommiddle)
    (inversemaskingb) {inverse masking};
    \node[block, mask] at (inversemaskingb |- inputmask-middleright) (masking) {masking};
    \draw pic
    at ($(preinput-northwest |- masking.south) + (0,-0.25in)$)
    (input) {NLBdata={}{mask}{spikes}{pointmask}};
    \node[block, below=1.125in of masking] (ANN) {ANN};
    \draw pic
    at ($(input-northwest |- ANN.south) + (0,-0.25in)$)
    (output) {NLBdata={INCLUDEOUT}{}{firingrates}{}};
    \draw pic at ($(inversemaskingb -| output-northwest) + (0,-0.5in)$)
    (maskedoutput) {NLBdata={INCLUDEOUT}{}{firingrates}{inversemask}};

    \draw pic[left=5.5in of preinput-topmiddle] (pretarget) {NLBdata={INCLUDEOUT}{}{spikes}{}};
    \node[block, mask] (inversemaskinga) at (outputmask-middleleft -| pretarget-bottommiddle) {inverse masking};
    \draw pic[below] at ($(inversemaskinga -| pretarget-northwest) + (0,-0.5in)$)
    (target) {NLBdata={INCLUDEOUT}{}{spikes}{inversemask}};

    \node[block] (loss) at (dynamicmask |- maskedoutput-middleleft) {loss};

    \draw[->,ultra thick] (trainingdata-bottommiddle) -- (split);
    \draw[->,ultra thick] (split) -- (preinput-middleleft);
    \draw[->,ultra thick] (split) -- ($(pretarget-middleright) + (0,0.07in)$);
    \draw[->,ultra thick] (pretarget-bottommiddle) -- (inversemaskinga);
    \draw[->,ultra thick] (preinput-bottommiddle) -- (masking);
    \draw[->,ultra thick] (masking) -- (input-topmiddle);
    \draw[->,ultra thick] (input-bottommiddle) -- (ANN);
    \draw[->,ultra thick] (ANN) -- (output-topmiddle);
    \draw[->,ultra thick] (output-bottommiddle) -- (inversemaskingb);
    \draw[->,ultra thick] (inversemaskingb) -- (maskedoutput-topmiddle);
    \draw[->,ultra thick] (inversemaskinga) -- (target-topmiddle);

    \draw[->,ultra thick] (dynamicmask) -- (inputmask-bottommiddle);
    \draw[->,ultra thick] (inputmask-middleright) -- (masking);
    \draw[->,ultra thick] (dynamicmask) -- (outputmask-topmiddle);
    \draw[->,ultra thick] (outputmask-middleleft) -- (inversemaskinga);
    \draw[->,ultra thick] (outputmask-middleright) -- (inversemaskingb);

    \draw[->,ultra thick] (target-middleright) -- (loss);
    \draw[->,ultra thick] (maskedoutput-middleleft) -- (loss);

\end{tikzpicture}%

}

\title{Inferring Population Dynamics in Macaque Cortex}
\author{Ganga Meghanath}
\author{Bryan Jimenez}
\author{Joseph G.\ Makin\footnote{Correspondence: \texttt{jgmakin@purdue.edu}. }}
\affil{%
    Elmore Family School of Electrical and Computer Engineering \authorcr
    Purdue University \authorcr
}

\begin{document}
\maketitle

\begin{abstract}
\emph{Objective.}
The proliferation of multi-unit cortical recordings over the last two decades, especially in macaques and during motor-control tasks, has generated interest in neural ``population dynamics'':\ the time evolution of neural activity across a group of neurons working together.
A good model of these dynamics should be able to infer the activity of unobserved neurons within the same population and of the observed neurons at future times.
Accordingly, Pandarinath and colleagues have introduced a benchmark to evaluate models on these two (and related) criteria:\ four data sets, each consisting of firing rates from a population of neurons, recorded from macaque cortex during movement-related tasks.
\emph{Approach.}
Since this is a discriminative-learning task, we hypothesize that general-purpose architectures based on recurrent neural networks (RNNs) trained with masking can outperform more ``bespoke'' models.
To capture long-distance dependencies without sacrificing the autoregressive bias of recurrent networks, we also propose a novel, hybrid architecture (``TERN'') that augments the RNN with self-attention, as in transformer networks.
\emph{Main results.}
Our RNNs outperform all published models on all four data sets in the benchmark.
The hybrid architecture improves performance further still.
Pure transformer models fail to achieve this level of performance, either in our work or that of other groups.
\emph{Significance.}
We argue that the autoregressive bias imposed by RNNs is critical for achieving the highest levels of performance, and establish the state of the art on the Neural Latents Benchmark.
We conclude, however, by proposing that the benchmark be augmented with an alternative evaluation of latent dynamics that favors generative over discriminative models like the ones we propose in this report.
\end{abstract}

\section*{Introduction} 
The primary motor cortex of rhesus macaques comprises tens of millions of neurons \cite{Herculano-Houzel2016}, so inferring the firing rates of any one neuron on the basis of a hundred others' might seem quixotic.
However, the state space of the body parts that motor cortex controls is much lower dimensional---on the order of hundreds of variables \cite{Prilutsky2002}.
And indeed, neural activity recorded by the Utah array \cite{Maynard1997}, which spans essentially the entire anteroposterior extent of the precentral gyrus, is correlated---at least during the reaching tasks typically employed in motor-control experiments \cite{Churchland2012,Sadtler2014}.

These observations have motivated the idea of ``population dynamics,'' i.e.\ a (possibly lossy) description of the dynamics of motor cortex as a whole, or at least those neurons picked up by a typical microelectrode array.
From this point of view, identifying population dynamics can be seen as a form of dimensionality reduction \cite{Churchland2012,Cunningham2014}.
More generally, the notion of a latent dynamical state maps naturally onto latent-variable generative models, and accordingly many attempts to capture neural dynamics have taken this form \cite{Yu2009,Aghagolzadeh2014,Aghagolzadeh2016a,Kao2015,Makin2018}.

Nevertheless, ground truth for evaluating models of population dynamics remains the ability to predict the activity either of the observed neurons, but at future times, or of neurons altogether unobserved.
Following \cite{Pei2021}, we refer to these as ``forward prediction'' and ``co-smoothing,'' respectively.
Still, adjudicating among different models has long been vexed by the use of proprietary data sets and idiosyncratic performance metrics.
Recently, the Neural Latents Benchmark (NLB) was introduced to solve these issues \cite{Pei2021}, proposing a single measure of performance---essentially the cross-entropy of a model that predicts ``instantaneous'' firing rates of unobserved neurons---and assembling four benchmark neural data sets. 
The data sets all consist of spiking activity recorded from single units with microelectrode arrays implanted into sensorimotor cortices (broadly construed) of rhesus macaques, who were performing motor or (in one case) cognitive tasks (see \sctname{methods}).

Here we show that standard recurrent neural networks (RNNs) achieve state-of-the-art performance on the benchmark.
Surprisingly, these RNNs outperform ``bespoke'' models designed with neural data in mind, as well as more recent neural-network architectures like transformers.
Furthermore, the RNN is not properly a latent-variable model; it is discriminative, rather than generative, and we train it essentially to map the observed to the held-out spiking activity.

We also show that the use of a transformer-encoder layer (i.e., with self-attention) \cite{Vaswani2017} can in fact improve performance over the ``standard'' RNN, but only when used after layers of RNN, rather than in the transformer architecture.
To our knowledge, this hybrid architecture is novel.
Pure transformer-based models, in contrast, whether in our hands or the hands of other groups \cite{Ye2021}, cannot achieve the level of performance of RNNs.
Altogether, these results suggest that, at least on data sets of the size found in the NLB, the autoregressive bias imposed by RNNs is worth the price of a less flexible model; but some of that flexibility can be usefully regained by augmenting an initially recurrent network with a layer of self-attention.

\section*{Results}\label{sec:results}
Our goal is to construct models that capture the underlying dynamics of a population of neurons.
Following the Neural Latents Benchmark (NLB) \cite{Pei2021}, neurons are taken to be a part of the same ``population'' if, roughly, all are located in the same Brodmann area and can be transduced with a single microelectrode array.
The neural dynamics have been ``captured'' if the model can infer the spiking activities of some neurons in the population based on contemporaneous recordings from others (``co-smoothing''); or again can predict the future spiking activity of the entire population of recorded neurons (``forward prediction'').
In particular, to quantify performance, each model is given a short
($\sim$500--1500-ms) 
sequence of spike counts across
$\sim$50--150 neurons,
and tasked with inferring the contemporaneous sequence of spike counts from a held-out set of neurons (about one third the size of the held-in set) simultaneously recorded by the array.
The same model is also tasked with predicting future spike-count sequences, for the unobserved as well as the observed neurons (see \sctname{methods}).

We trained all models identically---in short, to map the observed spike sequences to the unobserved spike sequences, both contemporaneous and future.
There is one wrinkle:\ about 25\% of the input bins were randomly set to zero (``masked'') on each full pass through the data, and the model was also required to predict the spike counts in these masked bins.
We explore the effect of this masking below.
The training procedure is described in detail in the \sctname{methods}.

\subsection*{Comparing \RNN, \transformer, and \TERN}
We consider three basic architectures (\fig{modelArchitectures}), the first two (\RNN, \transformer) being essentially standard time-series models in modern machine learning, albeit with a particular masking technique (see \sctname{methods}).
Our hypothesis is that state-of-the-art performance can be achieved with general-purpose architectures, not specifically designed for neural data.
However, the \RNN\ and \transformer\ make different assumptions about the data; or, more precisely, they impose different biases:
RNNs assume that temporal correlations in the data are mostly short-range, whereas the transformer is agnostic about which samples across the entire sequence are correlated with each other.
A single layer of an RNN can be thought of as a first-order, nonlinear, IIR filter---although we use gated units and multiple bidirectional layers, which allow for higher-order, noncausal filtering.
In contrast, a single layer of a transformer is essentially an arbitrarily powerful, non-causal, FIR filter.
(See \sctname{methods} for complete description of the models.)
Nevertheless, in theory, either model could, with sufficient data, learn to model any time series; but we are operating in a ``small-data'' regime, where there is a considerable trade-off between bias and variance (see also the \sctname{discussion} at the end of this report).

\FigArchitectures

As we show below, the \RNN\ consistently outperforms the \transformer, suggesting that most correlations in the data are short-range.
Still, we cannot rule out the existence of at least some long-range correlations, and accordingly we also investigate a third, hybrid architecture, consisting of an RNN (imposing a bias toward short-range dependencies), followed by a layer with self-attention (capable of finding long-range dependencies).
We call this model \TERN\ (Transformer Encoder over Recurrent Network; see \sctname{models} in the \sctname{methods} for full description of all three architectures).

    \begin{figure}[!t]
        \subfloat[][]{%
            \label{subfig:mc_maze_spikes_rates}%
            \scriptsize%
            \providecommand{\figheight}{1.0in}%
            \providecommand{\figwidth}{0.47\linewidth}%
            \input{\tikzdir/mc_maze_spikes_rates_ax2}
        }
        \subfloat[][]{%
            \label{subfig:mc_rtt_spikes_rates}%
            \scriptsize%
            \providecommand{\figheight}{1.0in}%
            \providecommand{\figwidth}{0.47\linewidth}%
            \input{\tikzdir/mc_rtt_spikes_rates_ax2}
        }\\
        \subfloat[][]{%
            \label{subfig:area2_bump_spikes_rates}%
            \scriptsize%
            \providecommand{\figheight}{1.0in}%
            \providecommand{\figwidth}{0.47\linewidth}%
            \input{\tikzdir/area2_bump_spikes_rates_ax2}
        }
        \subfloat[][]{%
            \label{subfig:dmfc_rsg_spikes_rates}%
            \scriptsize%
            \providecommand{\figheight}{1.0in}%
            \providecommand{\figwidth}{0.47\linewidth}%
            \input{\tikzdir/dmfc_rsg_spikes_rates_ax2}
        }\\
        \centering%
        \begin{minipage}[t]{0.32\textwidth}\vspace{0in}
        \subfloat[][]{%
            \label{subfig:mc_maze_velocity}%
            \scriptsize%
            \provideboolean{NOLEGEND}\setboolean{NOLEGEND}{false}%
            \provideboolean{CLEARYLABEL}\setboolean{CLEARYLABEL}{false}%
            \providecommand{\figheight}{1.1\linewidth}%
            \providecommand{\figwidth}{1.1\linewidth}%
            \input{\tikzdir/mc_maze_velocity}%
        }
        \end{minipage}
        \hspace{0.005\textwidth}
        \begin{minipage}[t]{0.32\textwidth}\vspace{0in}
        \subfloat[][]{%
            \label{subfig:mc_rtt_velocity}%
            \scriptsize%
            \provideboolean{NOLEGEND}\setboolean{NOLEGEND}{true}%
            \provideboolean{CLEARYLABEL}\setboolean{CLEARYLABEL}{true}%
            \providecommand{\figheight}{1.1\linewidth}%
            \providecommand{\figwidth}{1.1\linewidth}%
            \input{\tikzdir/mc_rtt_velocity}%
        }
        \end{minipage}
        \hspace{0.005\textwidth}
        \begin{minipage}[t]{0.32\textwidth}\vspace{0in}
        \subfloat[][]{%
            \label{subfig:area2_bump_velocity}%
            \scriptsize%
            \provideboolean{NOLEGEND}\setboolean{NOLEGEND}{true}%
            \provideboolean{CLEARYLABEL}\setboolean{CLEARYLABEL}{true}%
            \providecommand{\figheight}{1.1\linewidth}%
            \providecommand{\figwidth}{1.1\linewidth}%
            \input{\tikzdir/area2_bump_velocity}%
        }
        \end{minipage}
        \captioning{Example trials.}{%
        \subcaprange{mc_maze_spikes_rates}{dmfc_rsg_spikes_rates} Spike counts (in 5-ms bins) and firing-rate predictions for three randomly selected neurons (subplots) in each of the four data sets.
        The color of the lines for each firing-rate prediction indicates the model (green:\ \RNN, blue:\ \transformer, orange:\ \TERN).
        Note that the predictions have been converted from spikes/bin to spikes/s, shown on the left vertical axis.
        \subcaprange{mc_maze_velocity}{area2_bump_velocity}
        Velocity decoding (see \sctname{methods}) for \mcmaze, \mcrtt, and \areabump\ data set.  (\dmfcrsg\ does not have ground-truth hand velocity data due to the nature of the behavioral task.)}
        \label{fig:qualitativeResults}
    \end{figure}

\paragraph{Qualitative.}
\fig{qualitativeResults} shows qualitative results for selected trials.
In particular, \subfigss{mc_maze_spikes_rates}{dmfc_rsg_spikes_rates} illustrates decoding from trained instances of all three architectures, on hand-picked neurons and trials from the test partition of the data.
The models predict higher firing rates for regions with a high number of spikes and conversely, as expected.
More importantly, predictions from the \transformer\ are noticeably less smooth than the other two models, presumably because it lacks the autoregressive bias of RNNs.
A close inspection suggests \TERN\ is subtly smoother and more accurate even than the \RNN, but this needs to be quantified (see next).
(We also generate condition-averaged firing-rate predictions, i.e.\ model ``PSTHs''; see Supplementary \figss{mc_maze_PSTHs}{area2_bump_PSTHs}{dmfc_rsg_PSTHs}.)

\subfigss{mc_maze_velocity}{area2_bump_velocity} shows planar hand velocities that have been decoded (see \sctname{methods}) from the models' firing-rate predictions.
(The \dmfcrsg\ data set does not involve arm movements.)
More precisely, we fit a ridge regression on the training partition at each time step, mapping inferred firing rates for both held-in and held-out neurons to hand velocities.
The velocities and their predictions are plotted here for handpicked trials from the testing partition.
Qualitatively, we observed that the firing rates inferred from RNN-based models (\RNN, \TERN) produce smoother and more accurate trajectories than the \transformer.

\paragraph{Quantitative:\ co-bps, fp-bps, and velocity $R^2$.}
To quantify these apparent differences, we consider the 10 best instances (out of 125) of each architecture generated by our hyperparameter-optimization scheme (see \sctname{methods}).
Briefly, the ``best'' models are those that give the best inferences of unobserved firing rates, in terms of both co-smoothing and forward prediction.
These inferences are evaluated on a ``validation'' partition of each data set, whereas the results (\subfig{ourCosmoothing}, \fig{forwardPrediction} and \tbl{velocityPrediction}) are reported on a disjoint (local) test partition.

We consider first the ability of each model to infer the firing rates of unobserved neurons from the contemporaneous firing rates of observed neurons (co-smoothing \cite{Pei2021}).
Following the NLB, we ask how many bits per spike each model saves over a baseline model that simply predicts the average firing rates over thde whole trial (see \sctname{methods})---the so-called co-bps (\subfig{ourCosmoothing}).
We observe that the \RNN\ performs significantly better than \transformer\ on all four data sets.
We also note that training of the \transformer, although faster (as we expect, since within-layer computations can be executed in parallel in transformers), is less robust:\ performance across the top 10 instances varies much more for the \transformer\ than for the other two architectures.
In theory, this could indicate greater sensitivity to either hyperparameters or initialization of the weights; we suspect the latter (see below).
(We emphasize that our hyperparameter-optimization procedure was identical for all architectures.)

    \begin{figure}[!t]
        \subfloat[][]{%
            \label{subfig:ourCosmoothing}%
            \scriptsize%
            \newcommand{\figheight}{1.8in}%
            \newcommand{\figwidth}{0.283\linewidth}%
            \input{\tikzdir/co-bps_wilcoxon}%
        }\\
        \subfloat[][]{%
            \label{subfig:theirCosmoothing}%
            \scriptsize%
            \newcommand{\figwidth}{0.305\linewidth}%
            \newcommand{\figheight}{0.305\linewidth}%
            \input{\tikzdir/leaderboard}%
        }
        \captioning{Co-smoothing performance (co-bps).}{%
        The ability of each model to predict firing rates of contemporaneous, unobserved neurons, denominated in bits per spike, is shown separately for each of the four data sets (columns).
        \subcapref{ourCosmoothing}
        The distribution of co-bps across the 10 best models of each of our three architectures (colors as throughout).
        Boxes cover the interquartile ranges, with a line at the median; whiskers extend to the last datum within 150\% of the interquartile range from the end of the box.
        Any outliers beyond the whiskers are shown as circles.
        Significant differences between architectures, indicated with stars
        (*: $p<0.05$, **: $p<0.005$),
        were computed with a one-sided Wilcoxon rank-sum test, Holm–Bonferroni corrected for multiple comparisons (see \sctname{methods}).
        Results are from the ``local'' (publicly available) test set.
        \subcapref{theirCosmoothing}
        Comparison to published models on the Neural Latents Benchmark leaderboard.
        Numbers reported from \cite{Pei2021} and (when not available elsewhere) the leaderboard itself.
        For our models (\RNN, \transformer, and \TERN), we report the best of 10 submitted.
        Results are from the ``remote'' (not publicly available) test set.
        }
        \label{fig:cosmoothing}
    \end{figure}

\subfig{ourCosmoothing} also shows the co-smoothing rates for our hybrid RNN/transformer model, \TERN.
Like the pure RNN, it significantly outperforms the \transformer\ on all data sets.
Furthermore, \TERN\ is generally superior to \RNN; the advantage is statistically significant for the \mcmaze, \mcrtt, and \areabump\ datasets.
Since \TERN\ is essentially the \RNN\ plus a layer of a transformer encoder \cite{Vaswani2017} (see \fig{modelArchitectures}), it might seem that its performance advantage consists simply in having more parameters, but this is not the case:
The best instances of \TERN\ found by our hyperparameter optimization employ one or two layers of RNN, whereas the best \RNN\ models use two to four layers of RNN.

Next we turn to the models' forward predictions, i.e.\ their ability to predict firing rates of observed and unobserved neurons at time points beyond the last observed sample (\fig{forwardPrediction}).
Like co-smoothing, we quantify this ability in terms of the number of bits per spike (fp-bps) that each model saves over a baseline model that simply predicts that each neuron fire at its average rate over the entire set of future samples.
For two of the data sets (\mcmaze, \mcrtt), we find similar results:\ the RNN-based models are evidently superior to the \transformer\ for most time points, for both observed (top row) and unobserved neurons (bottom row).
For simplicity, we quantify this statistically only for the comparison of \TERN\ to the \transformer\ (broken red lines; \TERN\ is superior for 100\% [\mcmaze] and 95\% [\mcrtt] of time points for observed neurons and 85\% and 60\% for unobserved neurons).

%

For the other two data sets (\areabump, \dmfcrsg), \TERN\ is superior to the \transformer\ only about half the time in predicting observed neurons (50.0\% [\areabump] and 40.0\% [\dmfcrsg] of time points), and less still for unobserved neurons (27.5\%, 22.5\%).
We suspect that the \transformer\ is more competitive on forward prediction than co-smoothing because long-range dependencies are more useful for the former:\ a transformer can use the observations at the last sample to predict directly firing rates 40 samples away (the end of the forward-prediction task), whereas an RNN must propagate information from these observations through all 40 steps.
Nevertheless, this is sufficient to match performance of RNN-based architectures (\RNN\ and \TERN) only on data sets that lack strongly autocorrelated motor-control tasks:\ \areabump, in which the arm is unpredictably perturbed; and \dmfcrsg, a primarily cognitive task.

    \begin{figure}[!t]
        \centering
        \newcommand{\figwidth}{0.32\linewidth}%
        \newcommand{\figheight}{1.8in}%
        \scriptsize%
        \input{\tikzdir/fp-bps-time}
        \captioning{Forward-prediction performance (fp-bps-held-in and fp-bps-held-out).}{%
        The models' performances in predicting future firing rates of observed (top row) and unobserved (bottom row) neurons is plotted as a function of time---or, more precisely, samples---since the final observation.
        Solid lines and shaded regions indicate the mean (across the ten best model instances of each architecture) and its standard errors.
        The RNN/transformer hybrid \TERN\ outperforms the pure transformer at all points indicated by the (broken) red lines ($p<0.05$, Wilcoxon rank-sum test), computed separately at each time step.
        }\label{fig:forwardPrediction}
    \end{figure}

Next we consider velocity decoding (\tbl{velocityPrediction}).
As a baseline, we consider predictions made from the \emph{inputs}, rather than the outputs, of the models.
That is, we ask how well the spike counts of the held-in and held-out neurons can predict velocities through a linear (technically, affine) model.
Because the regression is fit on a sample-by-sample basis, predictions from raw spike counts are very poor (\tbl{velocityPrediction}, final column), even though the model has access to ground truth data.
Temporally smoothing the spike counts first \cite{Pei2021} (penultime column) greatly improves velocity decoding.
Nevertheless, the neural networks make much better predictions (second, third, fourth columns), even without access to the held-out neurons, because they incorporate information from multiple samples into a single estimate of the instantaneous firing rate at each time point.
Furthermore, and as expected from \subfigss{mc_maze_velocity}{area2_bump_velocity}, \TERN\ is superior to the \transformer\ on all data sets ($p < 0.0005$, Wilcoxon rank-sum test, Holm-Bonferroni corrected for two comparisons).
\TERN\ is likewise superior to the \RNN\ on \mcmaze\ and \mcrtt\ ($p < 0.005$), albeit by a smaller margin.
(Its apparent superiority to the \RNN\ on \areabump\ is not statistically significant.)

\TableVelocityPrediction

Finally, we note that during training of our models, the validation loss converged in fewer epochs for the \RNN\ than for the \transformer\ and \TERN\ (not shown).
Together, all these results strongly suggest that \emph{the bias imposed by recurrent neural networks toward short-term correlations is well matched to the population dynamics of neurons in monkey sensorimotor cortex.}
For the number of data in the Neural Latents Benchmark (see \tbl{databreakdown} in the \sctname{appendix}), this bias generates superior results to the more flexible transformer.
As we have seen, co-smoothing can be improved further still with the addition of a self-attention layer \emph{after} RNN layers, but not without the latter.

\subsection*{Comparison to other models}
In addition to the basic \RNN\ and \transformer\ architectures and our \TERN, many other models have been proposed for neural population dynamics, and tested on the Neural Latents Benchmark.
To compare our models against these, we submit them to the NLB website, where they are tested according to the same metrics as those employed in the previous section, but on a ``remote'' set of test data inaccessible to users except through this submit/test procedure.
(Performance can consequently differ by a small amount from the numbers reported above.)

\subfig{theirCosmoothing} shows co-smoothing scores for the best instance of each of our three architectures, along with the top published models on the NLB leaderboard.
(We have excluded ensemble models, including our own.)
To begin with, we note that our best \transformer\ models performed about the same as or better than the best Neural Data Transformer (NDT) constructed by another group, which suggests that our analyses apply to transformers more generally and not just our implementation.
(Where our instantiations are superior, we attribute it to better hyperparameter optimization.)


More importantly, our simple (GRU-based) \RNN\ evidently provides a better model of neuron population dynamics than the complex variants popular in the neuroscience literature, including the NDT, Gaussian-process factor analysis (GPFA) \cite{Yu2009} and the latest version of LFADS (latent factor analysis via dynamical systems, a variant on the variational autoencoder) \cite{Keshtkaran2022}.
This holds across all data sets except \dmfcrsg, the data set with a cognitive task, where performance is approximately matched by AutoLFADS and MINT \cite{Perkins2022}.
Adding post-RNN self-attention---i.e., the \TERN\ architecture---improves performance further still.
In fine, the \TERN\ architecture establishes state-of-the-art performance on the NLB, with the \RNN\ not far behind.
(Standard errors are available for some of these models from \cite{Pei2021}; see \tbl{leaderboard} in the \sctname{appendix}.)

\subsection*{Data-dependence of results}\label{sec:dataDependence}
The point of the Neural Latents Benchmark is to establish a uniform and general evaluation of models of neural dynamics \cite{Pei2021}.
Nevertheless, some facts about the benchmark data sets could be seen as arbitrary.
Here we consider how the results presented heretofore are affected by two such facts:\ the size of the data sets, and the resolution of the spike counts (5-ms bins).

\paragraph{The effect of reducing training data.}
We have seen that a model with high autoregressive bias, such as an RNN, is well suited for capturing neuron population dynamics, at least on the Neural Latents Benchmark---indeed, better than other architectures including the transformer.
But the competitiveness of the less-biased transformer is expected to increase with number of training data.
Accordingly, we examine how our results are affected by the number of data available for training.
We analyze the \mcmaze\ dataset, since it is the largest (most trials).
Keeping the test data the same as that of the original \mcmaze\ dataset, we optimized and trained all three architectures with four different subset sizes of the original dataset (150, 300, 600, and all $\sim$1250 trials).
The results are shown in \subfigss{mc_maze_scaled_co-bps}{mc_maze_scaled_velR2}.
As expected, performance generally increases with number of data for all architectures.
But the models with high autoregressive bias (\RNN\ and \TERN) are consistently significantly better than the \transformer\ in terms of co-bps, fp-bps, and velocity prediction, independent of the number of training data.
The gap between the more flexible transformer and the RNNs, which are biased toward short-range correlations, is expected to close at \emph{some} number of data (at the point where reducing bias is more important than reducing variance).
And indeed, the gap between the \transformer\ and the \RNN\ appears to have reduced by 1250 training trials.
But the gap between the \transformer\ and the hybrid \TERN\ mode does not appear to be narrowing over the range of data in even this, the largest NLB data set.

    \begin{figure}[!t]
        \centering%
        \newcommand{\figheight}{2.0in}
        \subfloat[][]{%
            \label{subfig:mc_maze_scaled_co-bps}%
            \begin{minipage}[t]{0.335\textwidth}
                \scriptsize%
                \newcommand{\figwidth}{1\linewidth}%
\begin{tikzpicture}
\provideboolean{CLEANXAXIS}\ifthenelse{\boolean{CLEANXAXIS}}{%
	\pgfplotsset{every axis post/.append style={xlabel = {} }}%
}{}%
\providecommand{\figwidth}{2in}%
\providecommand{\thisYlabelopacity}{1.0}%
\provideboolean{CLEANXAXIS}\ifthenelse{\boolean{CLEANXAXIS}}{%
	\pgfplotsset{every axis post/.append style={xticklabels = {} }}%
}{}%
\provideboolean{CLEANYAXIS}\ifthenelse{\boolean{CLEANYAXIS}}{%
	\pgfplotsset{every axis post/.append style={ylabel = {} }}%
}{}%
\provideboolean{CLEANYAXIS}\ifthenelse{\boolean{CLEANYAXIS}}{%
	\pgfplotsset{every axis post/.append style={yticklabels = {} }}%
}{}%
\providecommand{\figheight}{2in}%
\providecommand{\figwidth}{360pt}%
\providecommand{\figheight}{310pt}%
\provideboolean{CLEANTITLE}\ifthenelse{\boolean{CLEANTITLE}}{%
	\pgfplotsset{every axis post/.append style={title = {} }}%
}{}%
\provideboolean{NOLEGEND}%
\providecommand{\thisXlabelopacity}{1.0}%
\pgfplotsset{compat=1.15}%

\definecolor{cornflowerblue}{RGB}{100,149,237}
\definecolor{darkgray176}{RGB}{176,176,176}
\definecolor{forestgreen}{RGB}{34,139,34}
\definecolor{lightgray204}{RGB}{204,204,204}
\definecolor{orange}{RGB}{255,165,0}

\begin{axis}[
/pgf/number format/1000 sep={},
clip=false,
every axis x label/.append style={opacity=\thisXlabelopacity},
every axis y label/.append style={opacity=\thisYlabelopacity},
height=\figheight,
legend pos=south east,
legend style={fill opacity=0.8, draw opacity=1, text opacity=1, draw=lightgray204},
tick align=outside,
tick pos=left,
width=\figwidth,
x grid style={darkgray176},
x grid style={draw=black!15!white},
xlabel={no.\ training examples},
xmajorgrids,
xmin=92.95, xmax=1348.05,
xtick style={color=black},
xtick={ 150,300,600,1250 },
xticklabel style={rotate=90},
y grid style={darkgray176},
y grid style={draw=black!15!white},
ylabel={co-bps},
ymajorgrids,
ymin=0.208405788081466, ymax=0.391981385688714,
ytick style={color=black},
ytick={ 0.200,0.225,0.250,0.275,0.300,0.325,0.350,0.375,0.400 },
yticklabel style={/pgf/number format/fixed,/pgf/            number format/fixed zerofill,/pgf/number format/precision=3,}
]
\addplot [
  forget plot,
  mark=asterisk,
  only marks,
  scatter,
  scatter/@post marker code/.code={%
  \endscope
},
  scatter/@pre marker code/.code={%
  \expanded{%
  \noexpand\definecolor{thispointdrawcolor}{RGB}{\drawcolor}%
  \noexpand\definecolor{thispointfillcolor}{RGB}{\fillcolor}%
  }%
  \scope[draw=thispointdrawcolor, fill=thispointfillcolor]%
},
  visualization depends on={value \thisrow{draw} \as \drawcolor},
  visualization depends on={value \thisrow{fill} \as \fillcolor}
]
table{%
x  y  draw  fill
150 0.27593480172195 255,165,0 255,165,0
150 0.26593480172195 34,139,34 34,139,34
300 0.318926653657208 255,165,0 255,165,0
300 0.308926653657208 34,139,34 34,139,34
600 0.359184629434845 255,165,0 255,165,0
600 0.349184629434845 34,139,34 34,139,34
1291 0.38363704034293 255,165,0 255,165,0
1291 0.37363704034293 34,139,34 34,139,34
};
\path [draw=cornflowerblue, line width=2pt]
(axis cs:150,0.21675013342725)
--(axis cs:150,0.228083153294679);

\path [draw=cornflowerblue, line width=2pt]
(axis cs:300,0.269269456260108)
--(axis cs:300,0.281269999812747);

\path [draw=cornflowerblue, line width=2pt]
(axis cs:600,0.31343797817754)
--(axis cs:600,0.319055471346759);

\path [draw=cornflowerblue, line width=2pt]
(axis cs:1291,0.345846577192137)
--(axis cs:1291,0.352404292104145);

\addplot [semithick, cornflowerblue, mark=-, mark size=2.5, mark options={solid}, only marks, forget plot]
table {%
150 0.21675013342725
300 0.269269456260108
600 0.31343797817754
1291 0.345846577192137
};
\addplot [semithick, cornflowerblue, mark=-, mark size=2.5, mark options={solid}, only marks, forget plot]
table {%
150 0.228083153294679
300 0.281269999812747
600 0.319055471346759
1291 0.352404292104145
};
\path [draw=orange, line width=2pt]
(axis cs:150,0.235217797246033)
--(axis cs:150,0.246550817113462);

\path [draw=orange, line width=2pt]
(axis cs:300,0.291927405090226)
--(axis cs:300,0.303927948642865);

\path [draw=orange, line width=2pt]
(axis cs:600,0.332889137069009)
--(axis cs:600,0.338506630238228);

\path [draw=orange, line width=2pt]
(axis cs:1291,0.360358182886926)
--(axis cs:1291,0.366915897798934);

\addplot [semithick, orange, mark=-, mark size=2.5, mark options={solid}, only marks, forget plot]
table {%
150 0.235217797246033
300 0.291927405090226
600 0.332889137069009
1291 0.360358182886926
};
\addplot [semithick, orange, mark=-, mark size=2.5, mark options={solid}, only marks, forget plot]
table {%
150 0.246550817113462
300 0.303927948642865
600 0.338506630238228
1291 0.366915897798934
};
\path [draw=forestgreen, line width=2pt]
(axis cs:150,0.250268291788235)
--(axis cs:150,0.261601311655664);

\path [draw=forestgreen, line width=2pt]
(axis cs:300,0.292926381880888)
--(axis cs:300,0.304926925433527);

\path [draw=forestgreen, line width=2pt]
(axis cs:600,0.336375882850236)
--(axis cs:600,0.341993376019455);

\path [draw=forestgreen, line width=2pt]
(axis cs:1291,0.353280815452884)
--(axis cs:1291,0.359838530364892);

\addplot [semithick, forestgreen, mark=-, mark size=2.5, mark options={solid}, only marks, forget plot]
table {%
150 0.250268291788235
300 0.292926381880888
600 0.336375882850236
1291 0.353280815452884
};
\addplot [semithick, forestgreen, mark=-, mark size=2.5, mark options={solid}, only marks, forget plot]
table {%
150 0.261601311655664
300 0.304926925433527
600 0.341993376019455
1291 0.359838530364892
};
\addplot [semithick, cornflowerblue, forget plot]
table {%
150 0.222416643360965
300 0.275269728036428
600 0.31624672476215
1291 0.349125434648141
};
\addplot [semithick, orange, forget plot]
table {%
150 0.240884307179747
300 0.297927676866545
600 0.335697883653619
1291 0.36363704034293
};
\addplot [semithick, forestgreen, forget plot]
table {%
150 0.25593480172195
300 0.298926653657207
600 0.339184629434845
1291 0.356559672908888
};
\addlegendimage{line legend, cornflowerblue}\addlegendentry{\transformer}\addlegendimage{line legend, forestgreen}\addlegendentry{\RNN}\addlegendimage{line legend, orange}\addlegendentry{\TERN}
\ifthenelse{\boolean{NOLEGEND}}{\legend{}}{}
\end{axis}

\end{tikzpicture}
            \end{minipage}
        }
        \subfloat[][]{%
            \label{subfig:mc_maze_scaled_fp-bps}%
            \begin{minipage}[t]{0.335\textwidth}
                \vspace{-0.05in}%
                \scriptsize%
                \newcommand{\figwidth}{1\linewidth}%
\begin{tikzpicture}
\provideboolean{CLEANXAXIS}\ifthenelse{\boolean{CLEANXAXIS}}{%
	\pgfplotsset{every axis post/.append style={xlabel = {} }}%
}{}%
\providecommand{\figwidth}{2in}%
\providecommand{\thisYlabelopacity}{1.0}%
\provideboolean{CLEANXAXIS}\ifthenelse{\boolean{CLEANXAXIS}}{%
	\pgfplotsset{every axis post/.append style={xticklabels = {} }}%
}{}%
\provideboolean{CLEANYAXIS}\ifthenelse{\boolean{CLEANYAXIS}}{%
	\pgfplotsset{every axis post/.append style={ylabel = {} }}%
}{}%
\provideboolean{CLEANYAXIS}\ifthenelse{\boolean{CLEANYAXIS}}{%
	\pgfplotsset{every axis post/.append style={yticklabels = {} }}%
}{}%
\providecommand{\figheight}{2in}%
\providecommand{\figwidth}{360pt}%
\providecommand{\figheight}{310pt}%
\provideboolean{CLEANTITLE}\ifthenelse{\boolean{CLEANTITLE}}{%
	\pgfplotsset{every axis post/.append style={title = {} }}%
}{}%
\provideboolean{NOLEGEND}%
\providecommand{\thisXlabelopacity}{1.0}%
\pgfplotsset{compat=1.15}%

\definecolor{cornflowerblue}{RGB}{100,149,237}
\definecolor{darkgray176}{RGB}{176,176,176}
\definecolor{forestgreen}{RGB}{34,139,34}
\definecolor{lightgray204}{RGB}{204,204,204}
\definecolor{orange}{RGB}{255,165,0}

\begin{axis}[
/pgf/number format/1000 sep={},
clip=false,
every axis x label/.append style={opacity=\thisXlabelopacity},
every axis y label/.append style={opacity=\thisYlabelopacity},
height=\figheight,
legend pos=south east,
legend style={fill opacity=0.8, draw opacity=1, text opacity=1, draw=lightgray204},
tick align=outside,
tick pos=left,
width=\figwidth,
x grid style={darkgray176},
x grid style={draw=black!15!white},
xlabel={no.\ training examples},
xmajorgrids,
xmin=92.95, xmax=1348.05,
xtick style={color=black},
xtick={ 150,300,600,1250 },
xticklabel style={rotate=90},
y grid style={darkgray176},
y grid style={draw=black!15!white},
ylabel={fp-bps},
ymajorgrids,
ymin=0.127706326210024, ymax=0.263233042740952,
ytick style={color=black},
ytick={ 0.120,0.140,0.160,0.180,0.200,0.220,0.240,0.260,0.280 },
yticklabel style={/pgf/number format/fixed,/pgf/            number format/fixed zerofill,/pgf/number format/precision=3,}
]
\addplot [
  forget plot,
  mark=asterisk,
  only marks,
  scatter,
  scatter/@post marker code/.code={%
  \endscope
},
  scatter/@pre marker code/.code={%
  \expanded{%
  \noexpand\definecolor{thispointdrawcolor}{RGB}{\drawcolor}%
  \noexpand\definecolor{thispointfillcolor}{RGB}{\fillcolor}%
  }%
  \scope[draw=thispointdrawcolor, fill=thispointfillcolor]%
},
  visualization depends on={value \thisrow{draw} \as \drawcolor},
  visualization depends on={value \thisrow{fill} \as \fillcolor}
]
table{%
x  y  draw  fill
150 0.16785471550158 34,139,34 34,139,34
150 0.15785471550158 255,165,0 255,165,0
300 0.209405880733523 255,165,0 255,165,0
300 0.199405880733523 34,139,34 34,139,34
600 0.238318052475705 255,165,0 255,165,0
600 0.228318052475705 34,139,34 34,139,34
1291 0.257072737444092 255,165,0 255,165,0
1291 0.247072737444092 34,139,34 34,139,34
};
\path [draw=cornflowerblue, line width=2pt]
(axis cs:150,0.133866631506885)
--(axis cs:150,0.141827441973963);

\path [draw=cornflowerblue, line width=2pt]
(axis cs:300,0.17260721047906)
--(axis cs:300,0.17678336232979);

\path [draw=cornflowerblue, line width=2pt]
(axis cs:600,0.201379290596732)
--(axis cs:600,0.205438526426922);

\path [draw=cornflowerblue, line width=2pt]
(axis cs:1291,0.218109005981705)
--(axis cs:1291,0.224603017338662);

\addplot [semithick, cornflowerblue, mark=-, mark size=2.5, mark options={solid}, only marks, forget plot]
table {%
150 0.133866631506885
300 0.17260721047906
600 0.201379290596732
1291 0.218109005981705
};
\addplot [semithick, cornflowerblue, mark=-, mark size=2.5, mark options={solid}, only marks, forget plot]
table {%
150 0.141827441973963
300 0.17678336232979
600 0.205438526426922
1291 0.224603017338662
};
\path [draw=orange, line width=2pt]
(axis cs:150,0.141868403450559)
--(axis cs:150,0.149829213917638);

\path [draw=orange, line width=2pt]
(axis cs:300,0.187317804808159)
--(axis cs:300,0.191493956658888);

\path [draw=orange, line width=2pt]
(axis cs:600,0.215860205876388)
--(axis cs:600,0.219919441706578);

\path [draw=orange, line width=2pt]
(axis cs:1291,0.233825731765614)
--(axis cs:1291,0.240319743122571);

\addplot [semithick, orange, mark=-, mark size=2.5, mark options={solid}, only marks, forget plot]
table {%
150 0.141868403450559
300 0.187317804808159
600 0.215860205876388
1291 0.233825731765614
};
\addplot [semithick, orange, mark=-, mark size=2.5, mark options={solid}, only marks, forget plot]
table {%
150 0.149829213917638
300 0.191493956658888
600 0.219919441706578
1291 0.240319743122571
};
\path [draw=forestgreen, line width=2pt]
(axis cs:150,0.143874310268041)
--(axis cs:150,0.15183512073512);

\path [draw=forestgreen, line width=2pt]
(axis cs:300,0.186493549023208)
--(axis cs:300,0.190669700873938);

\path [draw=forestgreen, line width=2pt]
(axis cs:600,0.21628843456061)
--(axis cs:600,0.220347670390799);

\path [draw=forestgreen, line width=2pt]
(axis cs:1291,0.227785910386203)
--(axis cs:1291,0.23427992174316);

\addplot [semithick, forestgreen, mark=-, mark size=2.5, mark options={solid}, only marks, forget plot]
table {%
150 0.143874310268041
300 0.186493549023208
600 0.21628843456061
1291 0.227785910386203
};
\addplot [semithick, forestgreen, mark=-, mark size=2.5, mark options={solid}, only marks, forget plot]
table {%
150 0.15183512073512
300 0.190669700873938
600 0.220347670390799
1291 0.23427992174316
};
\addplot [semithick, cornflowerblue, forget plot]
table {%
150 0.137847036740424
300 0.174695286404425
600 0.203408908511827
1291 0.221356011660183
};
\addplot [semithick, orange, forget plot]
table {%
150 0.145848808684099
300 0.189405880733523
600 0.217889823791483
1291 0.237072737444092
};
\addplot [semithick, forestgreen, forget plot]
table {%
150 0.14785471550158
300 0.188581624948573
600 0.218318052475705
1291 0.231032916064681
};
\ifthenelse{\boolean{NOLEGEND}}{\legend{}}{}
\end{axis}

\end{tikzpicture}
            \end{minipage}
        }
        \subfloat[][]{%
            \label{subfig:mc_maze_scaled_velR2}%
            \begin{minipage}[t]{0.335\textwidth}
                \scriptsize%
                \newcommand{\figwidth}{1\linewidth}%
\begin{tikzpicture}
\provideboolean{CLEANXAXIS}\ifthenelse{\boolean{CLEANXAXIS}}{%
	\pgfplotsset{every axis post/.append style={xlabel = {} }}%
}{}%
\providecommand{\figwidth}{2in}%
\providecommand{\thisYlabelopacity}{1.0}%
\provideboolean{CLEANXAXIS}\ifthenelse{\boolean{CLEANXAXIS}}{%
	\pgfplotsset{every axis post/.append style={xticklabels = {} }}%
}{}%
\provideboolean{CLEANYAXIS}\ifthenelse{\boolean{CLEANYAXIS}}{%
	\pgfplotsset{every axis post/.append style={ylabel = {} }}%
}{}%
\provideboolean{CLEANYAXIS}\ifthenelse{\boolean{CLEANYAXIS}}{%
	\pgfplotsset{every axis post/.append style={yticklabels = {} }}%
}{}%
\providecommand{\figheight}{2in}%
\providecommand{\figwidth}{360pt}%
\providecommand{\figheight}{310pt}%
\provideboolean{CLEANTITLE}\ifthenelse{\boolean{CLEANTITLE}}{%
	\pgfplotsset{every axis post/.append style={title = {} }}%
}{}%
\provideboolean{NOLEGEND}%
\providecommand{\thisXlabelopacity}{1.0}%
\pgfplotsset{compat=1.15}%

\definecolor{cornflowerblue}{RGB}{100,149,237}
\definecolor{darkgray176}{RGB}{176,176,176}
\definecolor{forestgreen}{RGB}{34,139,34}
\definecolor{lightgray204}{RGB}{204,204,204}
\definecolor{orange}{RGB}{255,165,0}

\begin{axis}[
/pgf/number format/1000 sep={},
clip=false,
every axis x label/.append style={opacity=\thisXlabelopacity},
every axis y label/.append style={opacity=\thisYlabelopacity},
height=\figheight,
legend pos=south east,
legend style={fill opacity=0.8, draw opacity=1, text opacity=1, draw=lightgray204},
tick align=outside,
tick pos=left,
width=\figwidth,
x grid style={darkgray176},
x grid style={draw=black!15!white},
xlabel={no.\ training examples},
xmajorgrids,
xmin=92.95, xmax=1348.05,
xtick style={color=black},
xtick={ 150,300,600,1250 },
xticklabel style={rotate=90},
y grid style={darkgray176},
y grid style={draw=black!15!white},
ylabel={vel R2},
ymajorgrids,
ymin=0.776057398736347, ymax=0.937606309149166,
ytick style={color=black},
ytick={ 0.760,0.780,0.800,0.820,0.840,0.860,0.880,0.900,0.920,0.940 },
yticklabel style={/pgf/number format/fixed,/pgf/            number format/fixed zerofill,/pgf/number format/precision=3,}
]
\addplot [
  forget plot,
  mark=asterisk,
  only marks,
  scatter,
  scatter/@post marker code/.code={%
  \endscope
},
  scatter/@pre marker code/.code={%
  \expanded{%
  \noexpand\definecolor{thispointdrawcolor}{RGB}{\drawcolor}%
  \noexpand\definecolor{thispointfillcolor}{RGB}{\fillcolor}%
  }%
  \scope[draw=thispointdrawcolor, fill=thispointfillcolor]%
},
  visualization depends on={value \thisrow{draw} \as \drawcolor},
  visualization depends on={value \thisrow{fill} \as \fillcolor}
]
table{%
x  y  draw  fill
150 0.895214458803581 255,165,0 255,165,0
150 0.885214458803581 34,139,34 34,139,34
300 0.915985737819518 255,165,0 255,165,0
300 0.905985737819518 34,139,34 34,139,34
600 0.924457008583957 255,165,0 255,165,0
600 0.914457008583957 34,139,34 34,139,34
1291 0.930263176857674 255,165,0 255,165,0
1291 0.920263176857674 34,139,34 34,139,34
};
\path [draw=cornflowerblue, line width=2pt]
(axis cs:150,0.783400531027839)
--(axis cs:150,0.80038842522449);

\path [draw=cornflowerblue, line width=2pt]
(axis cs:300,0.839585312366616)
--(axis cs:300,0.852011389053527);

\path [draw=cornflowerblue, line width=2pt]
(axis cs:600,0.856162755902514)
--(axis cs:600,0.870655204578223);

\path [draw=cornflowerblue, line width=2pt]
(axis cs:1291,0.86823052621657)
--(axis cs:1291,0.879776689679176);

\addplot [semithick, cornflowerblue, mark=-, mark size=2.5, mark options={solid}, only marks, forget plot]
table {%
150 0.783400531027839
300 0.839585312366616
600 0.856162755902514
1291 0.86823052621657
};
\addplot [semithick, cornflowerblue, mark=-, mark size=2.5, mark options={solid}, only marks, forget plot]
table {%
150 0.80038842522449
300 0.852011389053527
600 0.870655204578223
1291 0.879776689679176
};
\path [draw=orange, line width=2pt]
(axis cs:150,0.855654711008886)
--(axis cs:150,0.872642605205537);

\path [draw=orange, line width=2pt]
(axis cs:300,0.879772699476063)
--(axis cs:300,0.892198776162974);

\path [draw=orange, line width=2pt]
(axis cs:600,0.887210784246103)
--(axis cs:600,0.901703232921811);

\path [draw=orange, line width=2pt]
(axis cs:1291,0.894490095126371)
--(axis cs:1291,0.906036258588977);

\addplot [semithick, orange, mark=-, mark size=2.5, mark options={solid}, only marks, forget plot]
table {%
150 0.855654711008886
300 0.879772699476063
600 0.887210784246103
1291 0.894490095126371
};
\addplot [semithick, orange, mark=-, mark size=2.5, mark options={solid}, only marks, forget plot]
table {%
150 0.872642605205537
300 0.892198776162974
600 0.901703232921811
1291 0.906036258588977
};
\path [draw=forestgreen, line width=2pt]
(axis cs:150,0.856720511705256)
--(axis cs:150,0.873708405901907);

\path [draw=forestgreen, line width=2pt]
(axis cs:300,0.874462015063223)
--(axis cs:300,0.886888091750134);

\path [draw=forestgreen, line width=2pt]
(axis cs:600,0.885919066885893)
--(axis cs:600,0.900411515561601);

\path [draw=forestgreen, line width=2pt]
(axis cs:1291,0.882866035189987)
--(axis cs:1291,0.894412198652593);

\addplot [semithick, forestgreen, mark=-, mark size=2.5, mark options={solid}, only marks, forget plot]
table {%
150 0.856720511705256
300 0.874462015063223
600 0.885919066885893
1291 0.882866035189987
};
\addplot [semithick, forestgreen, mark=-, mark size=2.5, mark options={solid}, only marks, forget plot]
table {%
150 0.873708405901907
300 0.886888091750134
600 0.900411515561601
1291 0.894412198652593
};
\addplot [semithick, cornflowerblue, forget plot]
table {%
150 0.791894478126164
300 0.845798350710072
600 0.863408980240368
1291 0.874003607947873
};
\addplot [semithick, orange, forget plot]
table {%
150 0.864148658107211
300 0.885985737819518
600 0.894457008583957
1291 0.900263176857674
};
\addplot [semithick, forestgreen, forget plot]
table {%
150 0.865214458803581
300 0.880675053406679
600 0.893165291223747
1291 0.88863911692129
};
\ifthenelse{\boolean{NOLEGEND}}{\legend{}}{}
\end{axis}

\end{tikzpicture}
            \end{minipage}
        }\\
        \subfloat[][]{%
            \label{subfig:binSizeEffects}%
            \scriptsize%
            \newcommand{\figwidth}{0.238\linewidth}%
            \renewcommand{\figheight}{1.7in}%
            \input{\tikzdir/SamplingEffect_mc_rtt}
        }
        \captioning{Data dependence of results.}{
        \subcaprange{mc_maze_scaled_co-bps}{mc_maze_scaled_velR2} Effect of number of training data (\mcmaze).
        For each of four training-set sizes and each of the three architectures, hyperparameters were optimized from scratch (see \sctname{methods}), and the top 10 (out of 125) models of each architecture selected.
        Mean and standard deviation of the \subcapref{mc_maze_scaled_co-bps} co-bps, \subcapref{mc_maze_scaled_fp-bps} fp-bps, and \subcapref{mc_maze_scaled_velR2} velocity $R^2$ are shown for each architecture.
        The \RNN\ and \TERN\ were tested (one-sided Wilcoxon rank-sum test with Holm-Bonferroni correction for two comparisons) for superiority to the \transformer; significance at $p<0.005$ is indicated with a star.
        \subcapref{binSizeEffects}
        Effect of spike-count resolution (\mcrtt).
        For each of the six bin sizes shown (columns) and the three architectures (abscissae), hyperparameters were optimized from scratch (see \sctname{methods}), and the top 10 (out of 125) models of each architecture selected.
        Mean co-bps (first row) and velocity $R^2$ (second row) are shown along with their standard deviations.
        Significance (*$: P<0.05$, **$: P<0.005$, ***$: P<0.0005$) is computed with a Wilcoxon rank-sum test, corrected for three comparisons.
        }\label{fig:dataffects}
    \end{figure}

\paragraph{The effect of bin size.}
For simplicity, we analyze only \mcrtt.
In particular, we re-sample the spike counts at 10, 15, 20, 25, and 30 ms (in addition to the original 5-ms bins), and then re-optimize models of all three architectures from scratch (see \sctname{methods}).
\subfig{binSizeEffects} shows the consequences for co-smoothing (co-bps, top row) and velocity predictions (coefficients of determination in bottom row).
In fine, the results are essentially unchanged:
The RNN-based models outperform the \transformer\ at all resolutions, and the hybrid \TERN\ is typically superior to the pure RNN.

\subsection*{The effect of masking}
A central contention of this report is that general-purpose RNNs, when properly optimized, perform as well---indeed, typically better---than more highly engineered models.
Admittedly, however, our RNN training procedure does contain one ``non-standard,'' albeit increasingly popular, component:\ dynamic masking of inputs \cite{Liu2020}.
Here we assess the impact of this component by training models with (1) the random point mask used throughout, (2) a mask of random ``strips'' of three consecutive time steps, and (3) no mask at all.
Note that in this last case, the output mispredictions are penalized at \emph{all} samples, including the $N_\text{in} \times M_\text{in}$ sub-array of held-in neurons; whereas with masking, the only held-in samples the model is asked to predict are those that have been masked out of the input (see \sctname{methods}).
Again for simplicity we focus on just \mcrtt\ and the \TERN\ architecture.
Evidently (Supplementary \fig{dynamicMaskingEffects}), dynamic masking significantly improves co-bps and velocity predictions (up from \transformer-level performance; cf.\ \tbl{velocityPrediction}).
The latter in particular is consistent with the fact that masking requires the model to ``fill in'' samples based on their neighbors, rather than copying them straight through.
Smoothing information across time is critical for velocity predictions since each is based on a single time point.

On the hypothesis that improved performance results from forcing the model to interpolate across time, we also tested the ``strip'' masks lately described, which require longer interpolations; but these resulted in performance very mildly (but significantly) inferior to the point mask (Supplementary \fig{dynamicMaskingEffects}, first and second bars).

\section*{Discussion}\label{sec:discussion}
\paragraph{RNNs provide state-of-the-art performance on the NLB.}
The notion of neural ``population dynamics'' arose as a consequence of multi-unit recordings:\ neural activity across scores of nearby ($\sim$mm) neurons was discovered to be highly correlated, at least during typical behavioral (especially motor-control) experiments.
One (but not the only) upshot is that it should be possible to ``fill in'' the contemporaneous activity of other nearby, but unobserved, neurons; and accordingly, the ability to do so was made the centerpiece of the Neural Latents Benchmark for modeling population dynamics.
The benchmark additionally includes metrics for predicting future neural activity and kinematic parameters (from the inferred, contemporaneous firing rates).

As stated, these are supervised-learning problems:\ map observed to unobserved firing rates.
Our investigation was premised on the hypothesis that general purpose ANNs, in particular RNNs with gated units, can provide high quality solutions to supervised-learning problems for time-series data, especially those that are dominated by short-range temporal correlations.
Indeed, this has already been shown for kinematic predictions \cite{Glaser2020}, albeit on a different data set.
In this report, we have verified the hypothesis:\ our RNNs match or outperform all other state-of-the-art architectures on the NLB (\subfig{theirCosmoothing}).

We note that one of the other models compared against in \subfig{theirCosmoothing} is likewise based on RNNs.
LFADS and its variants (in this case, AutoLFADS) use a pair of GRU-based RNNs to ``encode'' spike trains into a latent space and then decode them back into firing rates.
But LFADS is trained as a generative, not a discriminative, model.
In practice, generative models are most useful for discriminative tasks when extra, unlabelled data are available.
That is not the case in the NLB, which is based on a straightforward discriminative-learning task (map observed neural activity to unobserved neural activity); and consequently it seems likely that discriminative models, like ours, will yield the best performance.
On the other hand, the explicit latent space in generative models maps intuitively onto the notion of ``population dynamics''; we return to this point below.

\paragraph{Self-attention helps in small doses.}
In the last five years, transformers have eclipsed RNNs in popularity in several domains, largely on the strength of their performance in natural-language processing.
But the temporal dependency structure of natural language is not typical of time-series data in general:\ in the former, long-distance dependencies are common and there is no consistent short-term dependency structure.
(Natural images also contain important long-distance dependencies, because e.g.\ of edges.)
Transformers are ideal for such data because they make no assumptions about the order of their inputs.

In contrast, physical processes, and the time-series data describing them, typically do have consistent short-term dependencies (because they are governed by smooth differential equations).
This class includes the movements of the body but also (to a lesser degree) neural activity.
If the data were unlimited, these dependencies could be discovered by a transformer, and indeed the variance in parameter estimates incurred for its lack of bias could be driven down to arbitrarily small levels.
But the data are not unlimited, particularly in the NLB data sets.
Consequently, our RNNs consistently outperform transformers (from both our own and other groups) on the NLB (\figs{cosmoothing}{forwardPrediction}, \tbl{velocityPrediction}).

Nevertheless, self-attention can be helpful when the autoregressive bias is maintained.
Our \TERN\ architecture, which augments the RNN with a single layer of a transformer encoder, consistently outperforms the basic RNN (\figs{cosmoothing}{forwardPrediction}, \tbl{velocityPrediction}).
This model sets the state of the art for the benchmark.
To our knowledge it is novel, and we consider it to be one of the important contributions of this paper.
Indeed, it may have useful applications in other domains:\ wherever short-distance dependencies dominate, but long-distance correlations also exist.

Still, we ultimately expect experimental recording times to increase, especially as fully sensorized cages and wireless transmission make continuous, days- or weeks-long recording possible.
And for the largest data set (\mcmaze), our transformers appear to narrow the performance gap with the \RNN\ as the number of training data are increased (\subfig{dataSizeEffects}).
On the other hand, the gap with the \TERN\ does \emph{not} decrease.
Furthermore, the number of neurons recorded simultaneously will also increase, so perhaps the autoregressive bias will continue to be important for the foreseeable future.

\paragraph{Evaluating models of population dynamics.}
Finally, we might ask (in a more philosophical key) what it means to have modeled population dynamics.
As we have lately noted, the ability to infer unobserved neural activity is just one, and perhaps not the ideal, measure of the ability to capture latent dynamics, since it can be solved---indeed, at state-of-the-art levels---by discriminative models, as we have shown here.
Of course, a model that can predict the activity of held-out neurons on the basis of held-in ones must be extracting useful information from the latter!
But there are two possible objections:
(1) If a (discriminative) model can predict held-out activity while (say) completely ignoring some of the held-in neurons (or some aspect of their behavior, or etc.), it will.
(2) The discriminative model does not provide an unconditional description of the latent variables ($\generprior{} $).
It is possible in theory to assemble samples from this distribution (more precisely, from the ``aggregated posterior,'' i.e.\ $\recogposterior{} $ averaged under the data, $\datamarginal{} $); but our hidden layers have on the order of 100 units---too large of a space to be sampled with these small data sets.

Alternatively (or in addition), then, one might ask how well a model can reconstruct the \emph{observed} sequences of spikes ($\Heldins$), subject to a penalty on the size of the latent state ($\Generltnts$), to prevent the model from merely copying observations straight through to the output.
More precisely, one could evaluate the sum of the (average) reconstruction error (in bits or nats) and the (average) ``latent-code cost'' (also in bits),
\begin{equation*}
    \smplavg{\datamarginal{} \recogposterior{} }{-\log\generemission{patent=\Dataobsvs,latent=\Generltnts} }
    +
    \smplavg{\datamarginal{} \recogposterior{} }{-\log\generprior{latent=\Generltnts} }.
\end{equation*}
These average costs obviously depend on the distribution of observed spike sequences, $\datamarginal{} $.
But in generative models (like LFADS or GPFA), the ``recognition model''\cite{Dayan1995} or ``encoder'' \cite{Kingma2013} is stochastic, so the costs also depend on the probability, $\recogposterior{} $, with which the model assigns spike sequences to latent states $\Generltnts$.

A probabilistic encoder is a feature, not a bug, of such models---it allows for uncertainty about the latent state---and the model should not be penalized for it.
Indeed, if we were using the model for compression, we could in principle get these ``bits back'' \cite{Hinton1994} by employing some extra stream of data that we wish to communicate as the ``random'' seeds for draws from the recognition model.
That is, in this thought experiment, the ``sender'' would select a latent code vector $\generltnts$ for some input sequence $\dataobsvs$ by passing uniformly distributed \emph{data} (not noise) through the inverse cumulative distribution function (cdf) for $\recogposterior{patent=\dataobsvs} $.
He would pass the latent code along with the reconstruction error on to a receiver, at a cost of $-\log\generemission{patent=\dataobsvs,latent=\generltnts} - \log\generprior{latent=\generltnts} $ bits.
The receiver in turn could reconstruct $\dataobsvs$ from $\generltnts$ and the reconstruction error.
But she could also recover the extra stream of data by passing the latent codes back through the cdf of $\recogposterior{patent=\dataobsvs} $.
(Although this procedure was invented as a thought experiment \cite{Hinton1994}, some groups have recently attempted compression along these lines \cite{Townsend2019,Kingma2019}.)

This ``refund'' is worth $-\log\recogposterior{patent=\dataobsvs,latent=\generltnts} $ bits, so on average (across all observations and all assignments of latent codes) the total cost of the communicating information with the model is actually
\begin{equation*}
    \smplavg{\datamarginal{} \recogposterior{} }{-\log\generemission{patent=\Dataobsvs,latent=\Generltnts} }
    +
    \smplavg{\datamarginal{} \recogposterior{} }{-\log\generprior{latent=\Generltnts} }
    -
    \smplavg{\datamarginal{} \recogposterior{} }{-\log\recogposterior{patent=\Dataobsvs,latent=\Generltnts} } = \mathcal{F}.
\end{equation*}
This is precisely the ``free energy'' (a.k.a.\ the negative of the variational bound or ELBO), which is an upper bound on the model fit to the data.
Indeed, it is also precisely the loss that VAEs like LFADS are trained to minimize.
Although during training it is intended to be a surrogate for the marginal cross entropy, $\ntrp{\datadistrvar\generdistrvar}{\Dataobsvs;\params}$, that it upper bounds, the free energy itself is a more appropriate measure of the model's ability to capture latent dynamics:
The marginal cross entropy measures only the quality of the data generated by the model, taking no account of the inference/recognition/encoding mechanism.
The free energy measures both.
Of course, when the recognition model ($\recogposterior{} $) is equal to the posterior under the generative model ($\generposterior{} $), as for generative models that are tractably invertible with Bayes' rule, the free energy collapses into the marginal cross entropy.
But the free energy is more general.

It is perhaps surprising that, having designed and implemented models that outperform generative models on the benchmark, we conclude by arguing that the benchmark should be changed to be more favorable to the latter and less favorable to our own.
Indeed, even if one were to identify a hidden layer of a discriminative model as the ``latent variables,'' $\Generltnts$, it still would not be possible even to compute a free energy, since the models do not include a prior distribution, $\generprior{} $.
On the other hand, our models may nevertheless have discovered low-entropy (if not low-dimensional) latent codes for their inputs---although exploring this is beyond the scope of the present paper.
In any case, we think the free energy would make a useful addition to the NLB.

\section*{Methods}\label{sec:methods} 

\subsection*{Data sets}\label{sec:datasets}
The data consist of four sets of microelectrode array recordings from motor (broadly construed) or somatosensory cortex of rhesus macaques, made by different groups during different experiments, but collected and pre-processed identically for the Neural Latents Benchmark \cite{Pei2021} (see \fig{datasets} and \tbl{databreakdown} in the \sctname{appendix}).
We briefly describe the NLB pre-processing.
Raw voltages were spike sorted and binned at 5-ms resolution.
From these continuous recordings, snippets were extracted around each behavioral trial, e.g.\ by aligning to movement onset, yielding data arrays of size (number of neurons [$N$]) $\times$ (number of samples [$M$])---although note that $N$ and $M$ vary by data set.
Finally, approximately 25\% of all neurons and the final 40 samples (200 ms) of each trial were designated ``unobserved,'' thus dividing each trial's data array into four sub-arrays:\ held-in ($N_\text{in} \times M_\text{in}$), held-out ($N_\text{out} \times M_\text{in}$), held-in-forward ($N_\text{in} \times M_\text{fw}$) and held-out-forward ($N_\text{out} \times M_\text{fw}$), as shown in \fig{TrainingMethodology}.


\subsection*{Training} \label{sec:training}
The same training and evaluation pipeline, depicted in \fig{TrainingMethodology}, was used for all three architectures explored in this study (see \sctname{models} below).
Input consists of the array of observed spike counts from the held-in block ($N_\text{in} \times M_\text{in}$).
At each epoch, a random mask is drawn (\fig{TrainingMethodology}, center) and then used to mask (zero) out $\sim$27\% of the held-in spikes, similar to the Neural Data Transformer (NDT) \cite{Ye2021}.
The input data are then zero-padded out to the end of the entire trial (\fig{TrainingMethodology}, right), i.e.\ to size $N_\text{in} \times (M_\text{in} + M_\text{fw})$.
Note that the mask points for held-in spikes are generated independently across channels and time points; this is illustrated in \fig{TrainingMethodology}.

The output dimensions for all three models were set to
$(N_\text{in}+N_\text{out}) \times (M_\text{in}+M_\text{fw})$, i.e.\ spanning all four sub-arrays, and accordingly interpreted as predicted firing rates ($\vect{\lambda}$) for all neurons at all times, both observed and unobserved. 
However, the loss is computed only on those portions not available at the input:\ the masked elements of the held-in sub-array, as well as the entirety of the held-out, held-in-forward, and held-out-forward sub-arrays (\fig{TrainingMethodology}, bottom right).
(For the special case of no masking explored at the end of the \sctname{results}, the loss was computed on the entire held-in sub-array.)
For the held-in spikes, this is equivalent to the ``coordinated dropout'' that has been applied to LFADS models \cite{Keshtkaran2019}.

Since we aim to predict spike counts, a suitable loss is the cross entropy of the model under the data with respect to a Poisson distribution (see \sctname{evaluation}).
We reduce this loss by stochastic gradient descent (with AdaM optimization \cite{Kingma2014}) in the space of the model parameters, always using the maximal possible batch size on our GPUs.
(Typically this was one or two batches per epoch, but for the larger \mcmaze\ and \dmfcrsg\ data sets was 10 and 6, respectively.)
All models were trained on 60\% of the data (the training partition) and evaluated on a fixed 25\% (the test partition), with the remaining 15\% used for ``validation.''
That is, training was terminated (``early stopping'') and hyperparameters were chosen (see next) based on performance on this validation partition.

\paragraph{Hyperparameter optimization.}
The neural-network architectures we describe below depend on a number of hyperparameters, like number of layers and number of units per layer (see \tbl{hyperparameters} for full list).
We optimized these hyperparameters with the open-source Python package Optuna \cite{Akiba2019}, which efficiently searches the space with a mixture of kernel-density estimation (training runs that yield good results make the algorithm more likely to use nearby hyperparameters) and an evolutionary algorithm (that takes into account the relationships among hyperparameters).
We trained $\sim$125 model instances for each (architecture, dataset) pair, under the criterion of minimizing cross entropy on all three of the held-out, held-in-forward, and held-out-forward sub-arrays of the \emph{validation} partition.

When investigating how performance varies as a function of number of training data, spike-binning resolution, or masking (see \sctname{dataDependence}), we repeated this entire procedure at each of the data sizes, bin sizes, or mask choices.

\FigTraining

\subsection*{Evaluation}\label{sec:evaluation}
\paragraph{The objective.}
When evaluating the model (on the validation or test sets), the input is the same as during the training phase, except that no mask is applied.
That is, the input consists of the held-in sub-array ($N_\text{in} \times M_\text{in}$) (which is subsequently zero-padded out to size $N_\text{in} \times (M_\text{in} + M_\text{fw})$).
The model predicts firing rates for each neuron at all time steps (out to $M$) and all neurons (out to $N$), but following the NLB \cite{Pei2021}, we report performance separately on the three sub-arrays of unobserved data (see below).
Nevertheless, the performance criterion \cite{Pei2021} is the same for all three, a measure of information extracted per spike.
In particular, since the model outputs firing rates, $\rates$, it can be interpreted as assigning a conditional probability $\discrimdistrvar$ to each input-output pair according to a Poisson distribution:
\begin{equation*}
    \discrimdistribution{}
        \defeqleft
    \PoissonConditional{} ,
\end{equation*}
where $\params$ are the model parameters.
During training (see above), the parameters are changed so as to reduce the conditional cross entropy of this distribution with the observed distribution of input-output pairs, $\datamarginal{patent={\heldinsarg,\heldoutsarg}} $:
\begin{equation}\label{eqn:modelCrossEntropy}
    \begin{split}
        \ntrp{\datadistrvar\discrimdistrvar}{\Heldouts|\Heldins; \params}
            =
        \xpct{\datadistrvar}{-\log\discrimdistribution{inarg=\Heldins,outarg=\Heldouts} }
            &\approx
        \def\summand#1 {\log\discrimdistribution{#1} }
        \frac{-1}{N_\text{examples}}\samplessum{inarg/\heldins,outarg/\heldouts}{\summand}\\
            &=
        \def\summand#1 {\log\PoissonConditional{#1} }
        \frac{-1}{N_\text{examples}}\samplessum{inarg/\heldins,outarg/\heldouts}{\summand} ,
    \end{split}
\end{equation}
where the expectation under the data has been approximated with a sample average.
Intuitively, if a base-2 logarithm is used, this is the average number of yes/no questions (ideally, zero) one would have to ask in order to identify correctly all the unobserved spike counts, given the model predictions from the observed spike counts.

\paragraph{NLB metrics.}
The main two NLB evaluation metrics are identical to this cross entropy up to a shift and scaling.
In particular, a crude baseline for performance would be the cross entropy under a ``model'' that simply predicts that each neuron fire at its mean rate (over the unobserved samples), $\bar{\rates}$:
\begin{equation}\label{eqn:baselineCrossEntropy}
    \ntrp{\text{baseline}}{\Heldouts|\Heldins; \bar{\rates}}
        \approx
    \def\summand#1 {\log\BaselinePoissonConditional{params=\rates,#1} }
    \frac{-1}{N_\text{examples}}\samplessum{inarg/\heldins,outarg/\heldouts}{\summand}.
\end{equation}
We can then quantify model performance in terms of the improvement it provides over this baseline.
More precisely, the benchmark measures the reduction in bits (yes/no questions) that the model provides over the crude baseline, normalized by the total number of observed spikes:
\begin{equation}\label{eqn:bitsPerSpike}
    \text{bits}/\text{spike}
        =
    \left(
        \ntrp{\text{baseline}}{\Heldouts|\Heldins; \bar{\rates}} -
        \ntrp{\datadistrvar\discrimdistrvar}{\Heldouts|\Heldins; \params}
    \right)/N_\text{spikes}.
\end{equation}
(Since the ``baseline'' is constructed from the test partition, it could in theory outperform a real model.  In this, the NLB bits/spike resembles the coefficient of determination.)
Our training procedure, which minimizes the cross entropy in \eqn{modelCrossEntropy}, is clearly equivalent to maximizing the bits per spike in \eqn{bitsPerSpike}.

The benchmark proposes three main metrics:
\begin{itemize}
    \item{\textbf{co-bps}:
    \eqn{bitsPerSpike} applied to the ($N_\text{out} \times M_\text{in}$)-sized sub-array of held-out neurons, which measures how well the model can ``co-smooth'' (infer) contemporaneous spiking from unobserved neurons.
    In the NLB, this is considered the primary evaluation metric.
    }
    \item{\textbf{fp-bps}:
    \eqn{bitsPerSpike} applied to the ($N \times M_\text{fw}$)-sized sub-array of \emph{all} neurons in the 40 samples (200 ms) after the last observation, measuring how well the model can predict future activity of both observed and unobserved neurons (``forward prediction'').
    In practice, however, the models are (unsurprisingly) better at predicting the future activity of observed, rather than unobserved, neurons.
    To avoid obscuring this difference, we evaluate our models separately on fp-bps-held-out and fp-bps-held-in, i.e.\ the forward predictions on (respectively) the ($N_\text{in} \times M_\text{fw}$)-sized sub-array of observed neurons and the ($N_\text{out} \times M_\text{fw}$)-sized sub-array of held-out neurons.
    }
    \item{\textbf{velocity $R^2$}:
    the ability of the model's outputs to linearly predict the instantaneous velocities of the monkey's hand.
    The objective is to appraise the neural-prediction model, not the model that maps these to velocities, and accordingly the latter is supplied by the NLB and therefore uniform across all submissions.
    The linear model is fit (with ridge regression) on the model's responses to the training data (and the corresponding velocities), and evaluated in terms of coefficient of determination ($R^2$) on the testing partition.
    }
\end{itemize}

\paragraph{The test data.}
Most results reported here are on the ``local'' test partition, constructed by holding out 25\% of all trials.
However, the NLB also includes a second, ``remote'' test set that is not publicly available but on which models can be evaluated according to the performance metrics lately described, by submitting a model to the benchmark's website.
When comparing to architectures from other groups, we report these numbers (as we make clear in the \sctname{results}).
Models tested on the remote data were trained from scratch on the local test as well as training partitions (still reserving the validation set for early stopping), since the test partition would otherwise be wasted.

\subsection*{Model Architectures}\label{sec:models}
Guided by the hypothesis that standard machine-learning models, not inspired by or tailored specifically to neural data, can make competitive predictions on the Neural Latents Benchmark, we analyzed the three model architectures described here.
In contemporary machine learning, there are three dominant model paradigms for time-series data, all based on artificial neural networks (ANNs): recurrent neural networks (RNNs), transformers \cite{Vaswani2017}, and convolutional neural networks (CNNs).
We did not consider CNNs here and focused only on RNNs and transformers, including a hybrid RNN-transformer model, aiming to make use of the best features of both.
All instantiations of these three architectures (\fig{modelArchitectures}) were constructed, trained, and tested in PyTorch \cite{Paszke2017}.

\paragraph{\RNN.}
RNNs process their inputs sequentially, and consequently work better for data with shorter-length temporal autocorrelation.
The flipside of this virtue is that they struggle to learn sequences with long-distance correlations \cite{Cho2014a}.
This can be greatly ameliorated, although not eliminated, by introducing units that interact multiplicatively, rather than additively, allowing the model to learn to gate the flow of information and therefore preserve (and dump) information across long time scales \cite{Hochreiter1997}.
The RNNs investigated in the present study used gated recurrent units (GRUs) \cite{Cho2014b}; multiple recurrent layers; and a ``bidirectional'' architecture, i.e.\ each layer consisted of a pair of RNNs running in opposite temporal directions.
The final, purely feedforward layer used an exponential activation function to ensure that the predicted firing rates be positive.

\paragraph{\transformer.}
This model was inspired by the ``neural data transformer'' (NDT) introduced in \cite{Ye2021}, although our implementation is based on the RoBERTa formulation of the transformer \cite{Liu2020}, as implemented in the HuggingFace library.
It is essentially the encoder half of the transformer's encoder-decoder architecture \cite{Vaswani2017} (see \sctname{appendix} for hyperparameters).
In particular, any unit can transmit its activity at any time point to any unit in the next layer at any time point; but the flow of this information is governed by activity-dependent gating (``attention'').
We do not mask out future inputs, as in some implementations of the transformer, giving the network access to all neuron spikes at all (held-in) time steps.

Whereas in an RNN, the order of the input samples is implicitly coded by the order in which they enter the network, in an attention-based network this information is thrown away.
Accordingly, we encode position information with the sine/cosine method from the original transformer \cite{Vaswani2017}.
In our experiments, we place a feedforward layer (i.e., static affine transformation followed by pointwise nonlinearity) before and after the 
encoder.
The pointwise nonlinearity for the latter feedforward layer is exponential (like the \RNN) in order to predict positive firing rates.

\paragraph{\TERN.}
This (to our knowledge) novel architecture aims to combine the advantages of the \RNN\ and \transformer.
The model consists of an RNN followed by a single layer of a transformer encoder \cite{Vaswani2017}.
Our rationale is that the primary temporal correlations in neural data are local, and therefore an RNN provides a more appropriate model.
This is especially true for data sets such as \mcmaze\ and \mcrtt, where the (motor) neural activity coincides with a continuous reaching task:\ we expect some (although not all) of the kinematic correlations to be reflected in the neural commands and plans.
Nevertheless, we should like to allow the model to learn some long-distance correlations, after the short-distance correlations have been exploited.
Therefore we enable self-attention after the RNN, and accordingly call this model Transformer Encoder over Recurrent Network (TERN).

For consistency with the other two architectures, we use a bidirectional, GRU-based RNN, and a single layer of a transformer encoder \cite{Vaswani2017}.
After the fashion of RoBERTa \cite{Liu2020}, we do not use any attention mask for the transformer (i.e., we allow it to attend to the RNN's output at all time steps).
We also do not encode any position information between the RNN and Transformer layers. 

\subsection*{Significance testing}\label{sec:significanceTesting}
In order to quantify the statistical significance of architecture comparisons, we consider 10 model instances of each architecture.
Typically, for each architecture, one selects the set of hyperparameters that yielded best performance on the validation partition, and trains from scratch $N$ (in this case, 10) instances of that architecture with these hyperparameters.
Randomness in the initialization and the partition of training data into minibatches then yields a distribution of (non-hyper)parameterizations and consequently performances.

We consider this procedure somewhat brittle, since---at least in the present experiment---various settings of the hyperparameters yield fairly similar performance levels for a single architecture.
Consequently, the setting that yields the very best performance is not likely to generalize, especially since the validation partition on which the models were evaluated is arbitrary and not particularly large.

Therefore, for each architecture, we simply select the top 10 out of the 125 hyperparameter optimization runs (see above).
Since these 10 instances typically have different hyperparameters (numbers of layers, units, etc.), this means that in practice our results will be slightly noisier than under the more standard procedure, and significance slightly harder to prove.
Nevertheless, we believe this increase in variance is worth the reduction in bias, making the results more reliably informative about the different architectures.
We could reduce variance as well by considering more than 10 instances of each architecture, but informally we notice that models outside of the top 10 can be far from optimal.

It is difficult to justify an assumption of Gaussianity for a distribution of just 10 data, so we do not use $t$ tests for our comparisons.
Instead, we use the Wilcoxon rank-sum test (a.k.a.\ the Mann-Whitney $U$ test), which essentially asks whether the probability mass of one distribution (of $X$) is shiftward upward with respect to another distribution (of $Y$).
More precisely, it tests the null hypothesis that the probability that $X > Y$ is the same as the probability that $X < Y$.
Since we are hypothesizing that \TERN\ is superior to the \RNN, and that both are superior to the \transformer, we use single-sided tests throughout.
The exception is in our comparison of models with different types of masks:\ here we are agnostic as to which is superior and accordingly use a two-sided test.
When comparing multiple pairs of architectures with respect to the same hypothesis (e.g., co-bps), we correct the $p$ values with the Holm-Bonferroni method.

\section*{Data availability}
All data are from the Neural Latents Benchmark \cite{Pei2021} and freely available at their website.

\section*{Code availability}
All code will be made available via github at publication.

\section*{Acknowledgments}
This work was funded by a seed grant from the Brain Research Foundation.
Some neural networks were trained a GPUs generously donated by the Nvidia corporation.

\section*{Author contributions}
G.M.\ and B.J.\ designed, implemented, and tested all the models; BJ devised the \TERN\ architecture.
J.G.M.\ and G.M.\ wrote the paper.
J.G.M.\ supervised and advised the project.

\bibliographystyle{plain}
\bibliography{%
    ../../bibs/neuroscience,%
    ../../bibs/compneuro,%
    ../../bibs/machinelearning,%
    ../../bibs/BMI,%
    ../../bibs/nonpapers,%
    ../../bibs/misctech,%
}

\clearpage
\setcounter{page}{1}
\appendix
\section{Supplementary Material}\label{sec:appendix}
\setcounter{figure}{0}
\setcounter{table}{0}
\renewcommand{\figurename}{Supplementary Figure}
\renewcommand{\tablename}{Supplementary Table}

    \begin{figure}[!h]
        \scriptsize%
        \newlength\FullTextWidth%
        \setlength\FullTextWidth\textwidth%
        \newcommand{\pngwidth}{2.2}%
        \newcommand{\mcmazewidth}{4}%
        \newcommand{\areabumpwidth}{4}%
        \newcommand{\mcrttwidth}{2.5}%
        \newcommand{\dmfcwidth}{6}%
        \newcommand{\spacewidth}{1}%
        \pgfmathsetmacro{\denominator}{2*\pngwidth + \mcmazewidth + \areabumpwidth + \spacewidth}%
        \pgfmathparse{(\pngwidth + \mcmazewidth)/\denominator}%
        \begin{minipage}[c]{\pgfmathresult\FullTextWidth}\vspace{0in}
            \subfloat[][]{%
                \label{subfig:mc_maze}
                \begin{tabular}[b]{c}
                    \mcmaze: Dorsal premotor cortex + primary motor cortex\\
                    \pgfmathparse{\pngwidth/\denominator}%
                    \begin{minipage}[t]{\pgfmathresult\FullTextWidth}\vspace{0in}
                        \includegraphics[width=\linewidth]{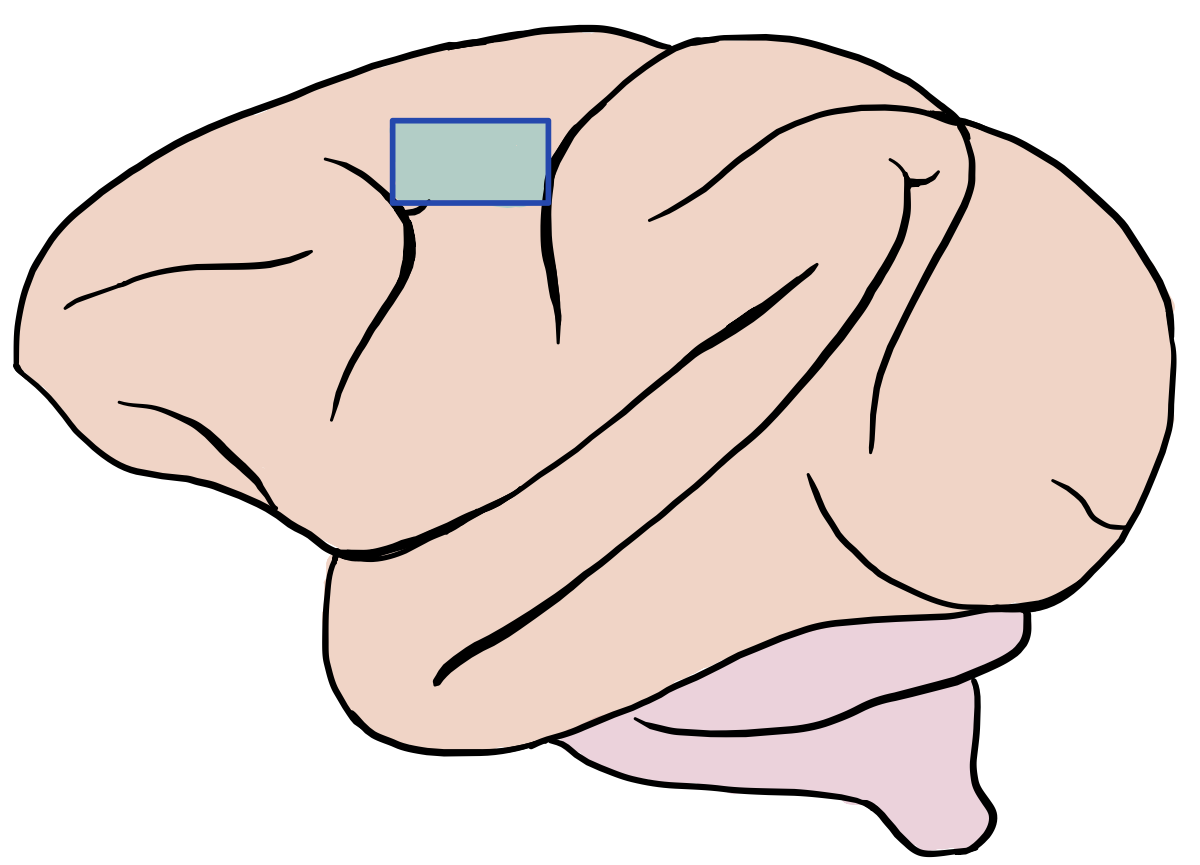}\\
                        \includegraphics[width=\linewidth]{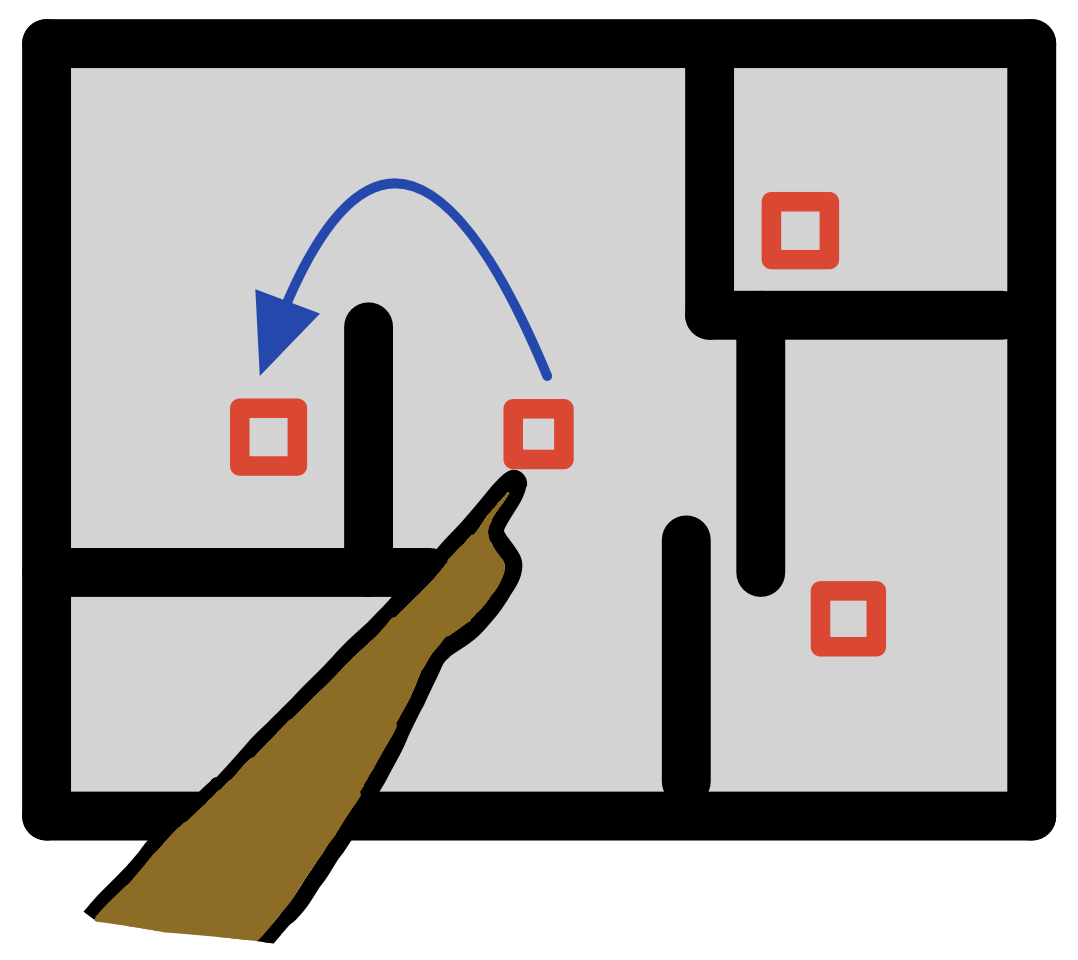}
                    \end{minipage}
                    \pgfmathparse{\mcmazewidth/\denominator}%
                    \begin{minipage}[t]{\pgfmathresult\FullTextWidth}\vspace{0in}
                        \scriptsize%
                        \providecommand{\figheight}{2in}%
                        \providecommand{\figwidth}{\linewidth}%
\begin{tikzpicture}
\providecommand{\thisXlabelopacity}{1.0}%
\providecommand{\figwidth}{360pt}%
\pgfplotsset{compat=1.15}%
\provideboolean{CLEARXLABEL}\ifthenelse{\boolean{CLEARXLABEL}}{%
	\pgfplotsset{every axis post/.append style={xlabel = {} }}%
}{}%
\provideboolean{CLEANYAXIS}\ifthenelse{\boolean{CLEANYAXIS}}{%
	\pgfplotsset{every axis post/.append style={yticklabels = {} }}%
}{}%
\provideboolean{CLEANTITLE}\ifthenelse{\boolean{CLEANTITLE}}{%
	\pgfplotsset{every axis post/.append style={title = {} }}%
}{}%
\provideboolean{CLEARYLABEL}\ifthenelse{\boolean{CLEARYLABEL}}{%
	\pgfplotsset{every axis post/.append style={ylabel = {} }}%
}{}%
\provideboolean{NOLEGEND}%
\providecommand{\thisYlabelopacity}{1.0}%
\provideboolean{CLEANXAXIS}\ifthenelse{\boolean{CLEANXAXIS}}{%
	\pgfplotsset{every axis post/.append style={xticklabels = {} }}%
}{}%
\provideboolean{CLEANYAXIS}\ifthenelse{\boolean{CLEANYAXIS}}{%
	\pgfplotsset{every axis post/.append style={ylabel = {} }}%
}{}%
\providecommand{\figheight}{310pt}%
\provideboolean{CLEANXAXIS}\ifthenelse{\boolean{CLEANXAXIS}}{%
	\pgfplotsset{every axis post/.append style={xlabel = {} }}%
}{}%

\definecolor{chocolate217952}{RGB}{217,95,2}
\definecolor{darkgray176}{RGB}{176,176,176}

\begin{axis}[
clip=false,
every axis x label/.append style={opacity=\thisXlabelopacity},
every axis y label/.append style={opacity=\thisYlabelopacity},
height=\figheight,
tick align=outside,
tick pos=left,
width=\figwidth,
x grid style={darkgray176},
xlabel={spikes/bin},
xmajorgrids,
xmin=-0.59, xmax=3.59,
xtick style={color=black},
xtick={0.0,1.0,2.0,3.0},
y grid style={darkgray176},
ylabel={\(\displaystyle \log_2\)(count)},
ymajorgrids,
ymin=0, ymax=27.008296869892,
ytick style={color=black}
]
\draw[draw=none,fill=chocolate217952,fill opacity=0.9] (axis cs:-0.4,0) rectangle (axis cs:0.4,25.7221874951352);
\draw[draw=none,fill=chocolate217952,fill opacity=0.9] (axis cs:0.6,0) rectangle (axis cs:1.4,19.955191557943);
\draw[draw=none,fill=chocolate217952,fill opacity=0.9] (axis cs:1.6,0) rectangle (axis cs:2.4,13.1819286925126);
\draw[draw=none,fill=chocolate217952,fill opacity=0.9] (axis cs:2.6,0) rectangle (axis cs:3.4,5.04439411935845);
\ifthenelse{\boolean{NOLEGEND}}{\legend{}}{}
\end{axis}

\end{tikzpicture}%
                    \end{minipage}
                \end{tabular}
            }
        \end{minipage}
        \hfill
        \pgfmathparse{(\pngwidth + \areabumpwidth)/\denominator}%
        \begin{minipage}[c]{\pgfmathresult\FullTextWidth}\vspace{0in}
            \subfloat[][]{%
                \label{subfig:area2bump}%
                \begin{tabular}[b]{c}
                    \areabump: Area 2 (somatosensory cortex)\\
                    \pgfmathparse{\pngwidth/\denominator}%
                    \begin{minipage}[t]{\pgfmathresult\FullTextWidth}\vspace{0in}
                        \includegraphics[width=\linewidth]{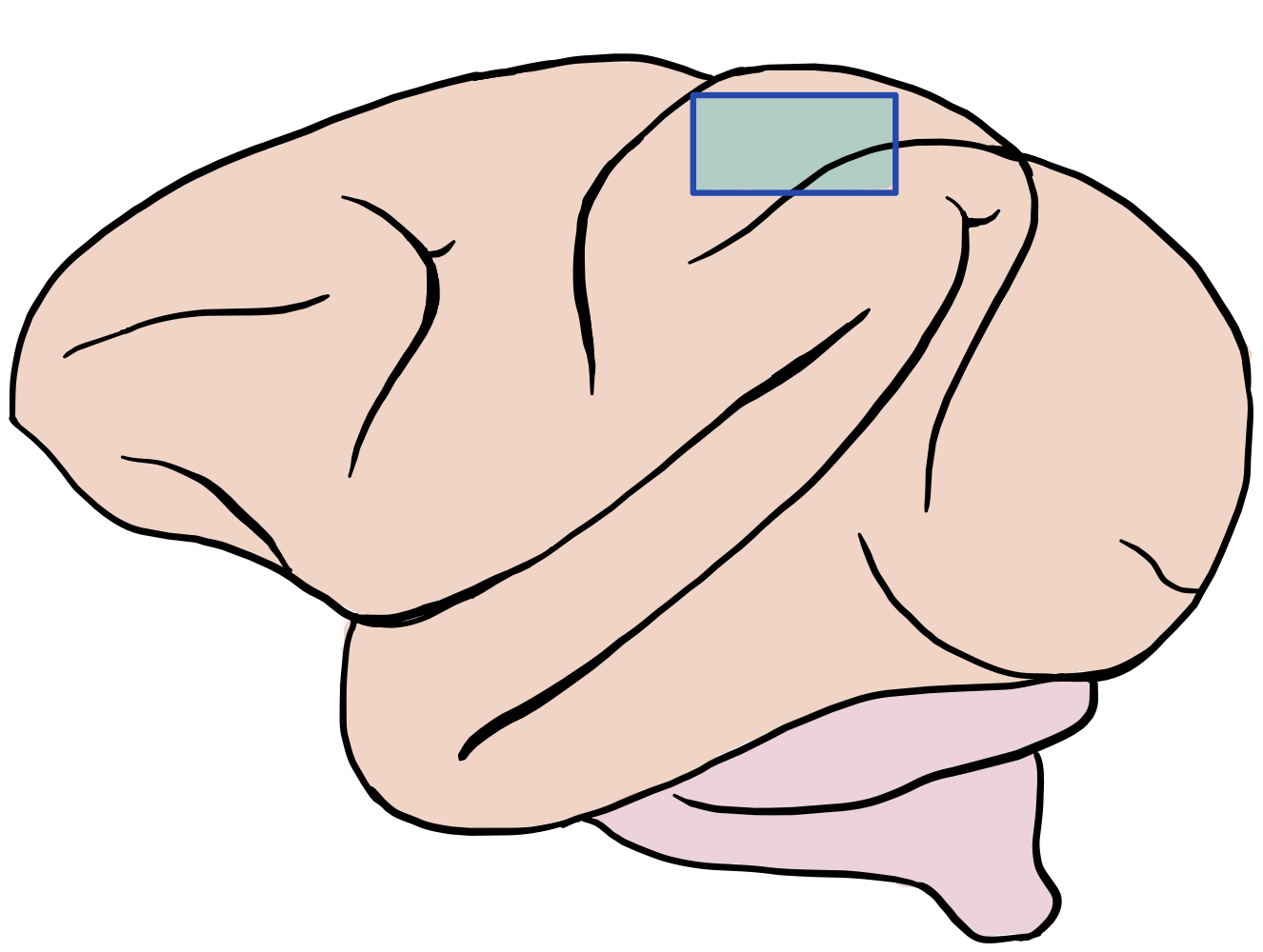}\\
                        \includegraphics[width=\linewidth]{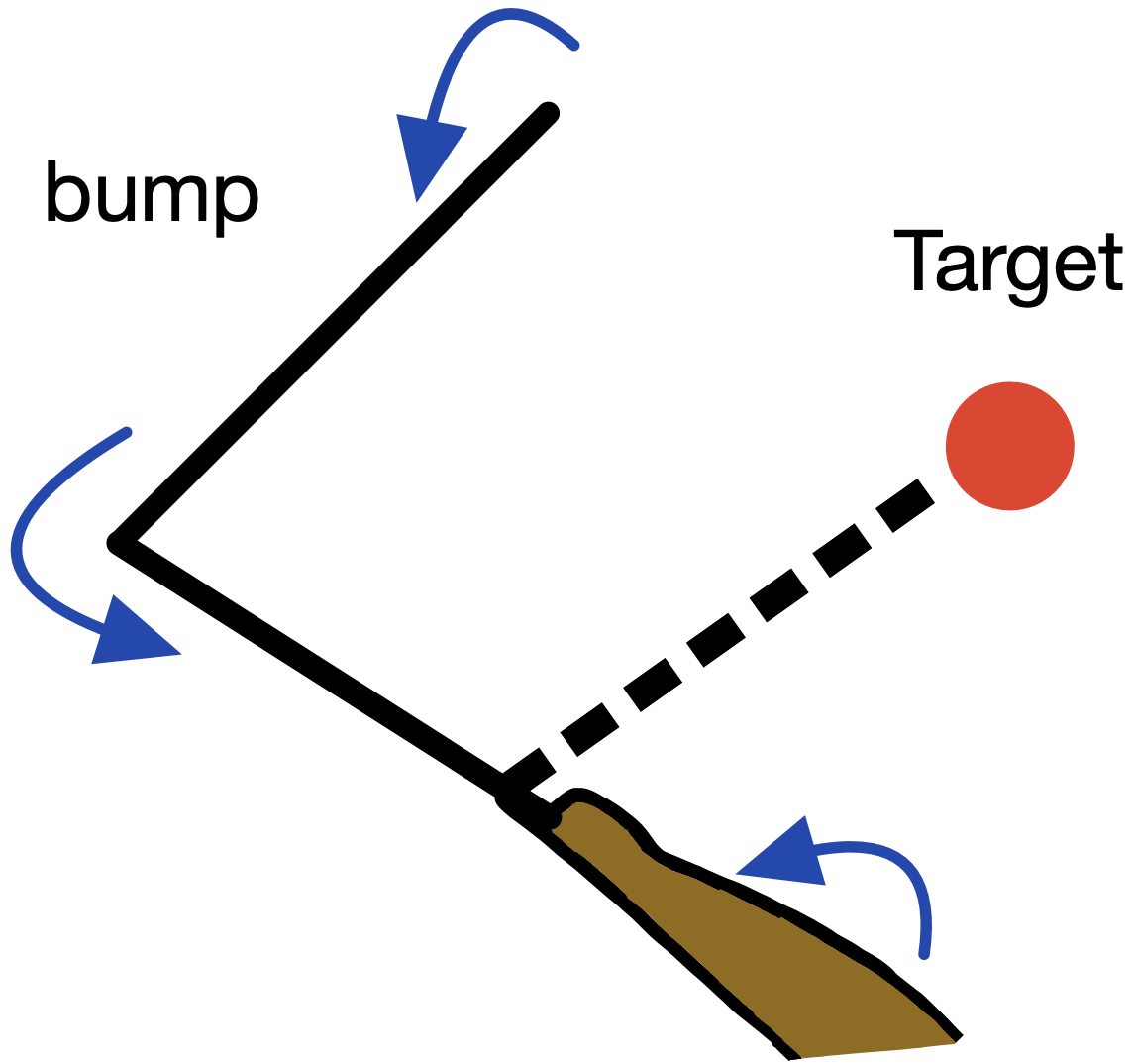}
                    \end{minipage}
                    \pgfmathparse{\areabumpwidth/\denominator}
                    \begin{minipage}[t]{\pgfmathresult\FullTextWidth}\vspace{0in}
                        \scriptsize%
                        \providecommand{\figheight}{2in}%
                        \providecommand{\figwidth}{\linewidth}%
\begin{tikzpicture}
\providecommand{\thisXlabelopacity}{1.0}%
\providecommand{\figwidth}{360pt}%
\pgfplotsset{compat=1.15}%
\provideboolean{CLEARXLABEL}\ifthenelse{\boolean{CLEARXLABEL}}{%
	\pgfplotsset{every axis post/.append style={xlabel = {} }}%
}{}%
\provideboolean{CLEANYAXIS}\ifthenelse{\boolean{CLEANYAXIS}}{%
	\pgfplotsset{every axis post/.append style={yticklabels = {} }}%
}{}%
\provideboolean{CLEANTITLE}\ifthenelse{\boolean{CLEANTITLE}}{%
	\pgfplotsset{every axis post/.append style={title = {} }}%
}{}%
\provideboolean{CLEARYLABEL}\ifthenelse{\boolean{CLEARYLABEL}}{%
	\pgfplotsset{every axis post/.append style={ylabel = {} }}%
}{}%
\provideboolean{NOLEGEND}%
\providecommand{\thisYlabelopacity}{1.0}%
\provideboolean{CLEANXAXIS}\ifthenelse{\boolean{CLEANXAXIS}}{%
	\pgfplotsset{every axis post/.append style={xticklabels = {} }}%
}{}%
\provideboolean{CLEANYAXIS}\ifthenelse{\boolean{CLEANYAXIS}}{%
	\pgfplotsset{every axis post/.append style={ylabel = {} }}%
}{}%
\providecommand{\figheight}{310pt}%
\provideboolean{CLEANXAXIS}\ifthenelse{\boolean{CLEANXAXIS}}{%
	\pgfplotsset{every axis post/.append style={xlabel = {} }}%
}{}%

\definecolor{darkgray176}{RGB}{176,176,176}
\definecolor{lightslategray117112179}{RGB}{117,112,179}

\begin{axis}[
clip=false,
every axis x label/.append style={opacity=\thisXlabelopacity},
every axis y label/.append style={opacity=\thisYlabelopacity},
height=\figheight,
tick align=outside,
tick pos=left,
width=\figwidth,
x grid style={darkgray176},
xlabel={spikes/bin},
xmajorgrids,
xmin=-0.59, xmax=3.59,
xtick style={color=black},
xtick={0.0,1.0,2.0,3.0},
y grid style={darkgray176},
ylabel={\(\displaystyle \log_2\)(count)},
ymajorgrids,
ymin=0, ymax=22.4398897846183,
ytick style={color=black}
]
\draw[draw=none,fill=lightslategray117112179,fill opacity=0.9] (axis cs:-0.4,0) rectangle (axis cs:0.4,21.3713236043984);
\draw[draw=none,fill=lightslategray117112179,fill opacity=0.9] (axis cs:0.6,0) rectangle (axis cs:1.4,16.7942889344713);
\draw[draw=none,fill=lightslategray117112179,fill opacity=0.9] (axis cs:1.6,0) rectangle (axis cs:2.4,11.2179576978641);
\draw[draw=none,fill=lightslategray117112179,fill opacity=0.9] (axis cs:2.6,0) rectangle (axis cs:3.4,4);
\ifthenelse{\boolean{NOLEGEND}}{\legend{}}{}
\end{axis}

\end{tikzpicture}
                    \end{minipage}
                \end{tabular}
            }
        \end{minipage}
        \pgfmathsetmacro{\denominator}{2*\pngwidth + \mcrttwidth + \dmfcwidth + \spacewidth}%
        \pgfmathparse{(\pngwidth + \mcrttwidth)/\denominator}%
        \begin{minipage}[c]{\pgfmathresult\FullTextWidth}\vspace{0in}
            \subfloat[][]{%
                \label{subfig:mc_rtt}%
                \begin{tabular}[b]{c}
                    \mcrtt: Primary motor cortex\\
                    \pgfmathparse{\pngwidth/\denominator}%
                    \begin{minipage}[t]{\pgfmathresult\FullTextWidth}\vspace{0in}
                        
                        \includegraphics[width=\linewidth]{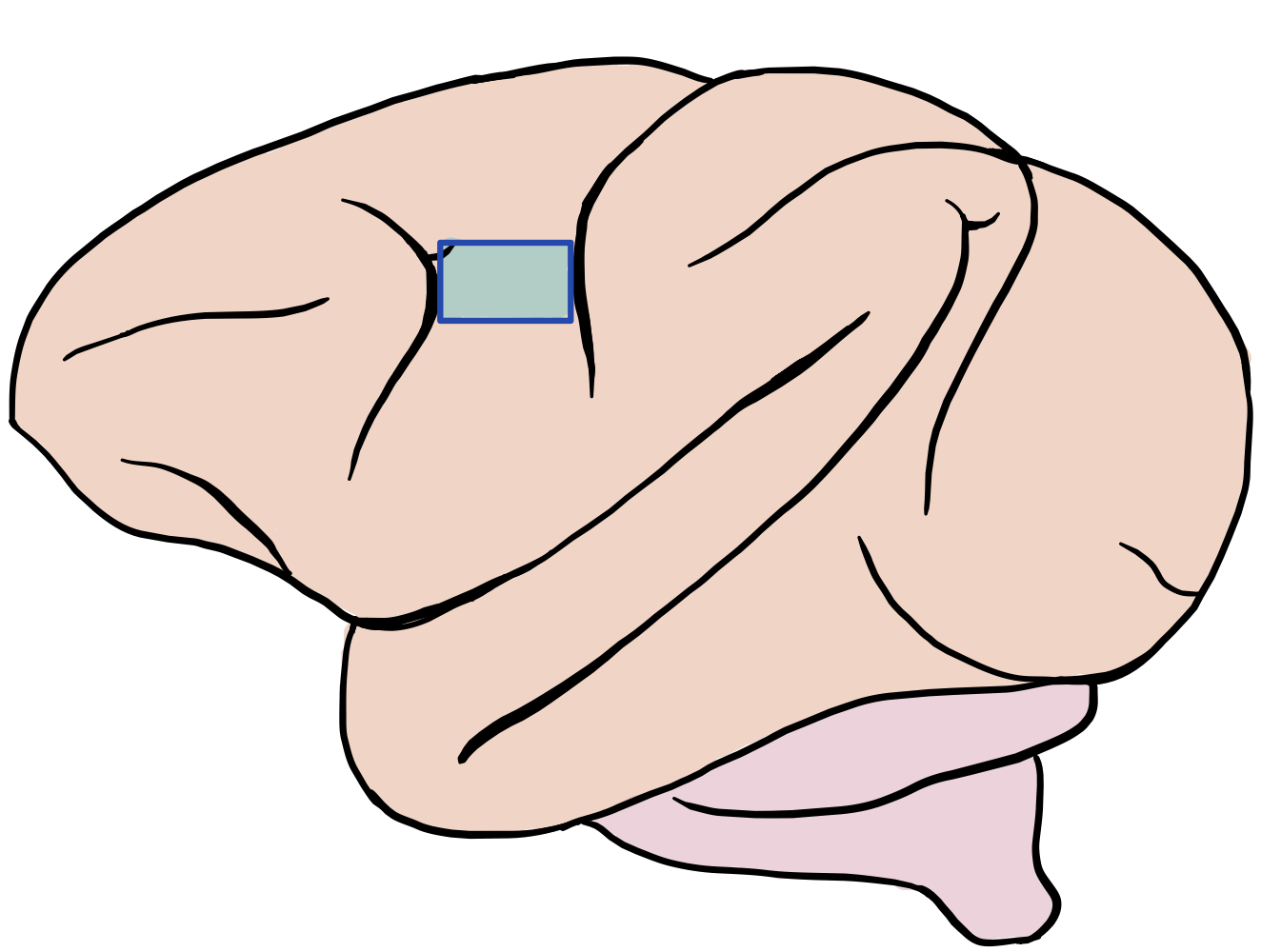}\\
                        \includegraphics[width=\linewidth]{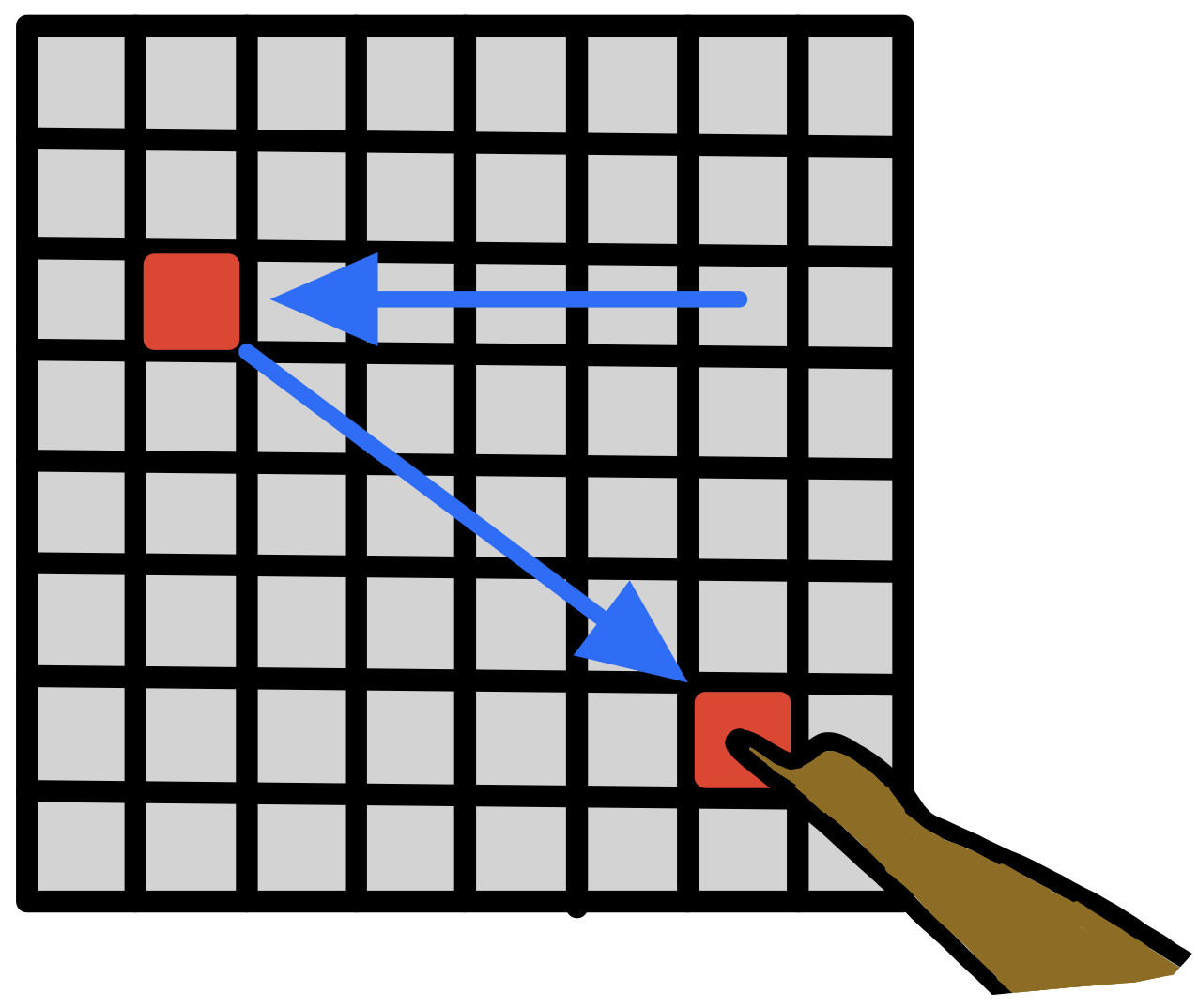}    
                    \end{minipage}
                    \pgfmathparse{\mcrttwidth/\denominator}
                    \begin{minipage}[t]{\pgfmathresult\FullTextWidth}\vspace{0in}
                        \scriptsize%
                        \providecommand{\figheight}{2in}%
                        \providecommand{\figwidth}{\linewidth}%
\begin{tikzpicture}
\providecommand{\thisXlabelopacity}{1.0}%
\providecommand{\figwidth}{360pt}%
\pgfplotsset{compat=1.15}%
\provideboolean{CLEARXLABEL}\ifthenelse{\boolean{CLEARXLABEL}}{%
	\pgfplotsset{every axis post/.append style={xlabel = {} }}%
}{}%
\provideboolean{CLEANYAXIS}\ifthenelse{\boolean{CLEANYAXIS}}{%
	\pgfplotsset{every axis post/.append style={yticklabels = {} }}%
}{}%
\provideboolean{CLEANTITLE}\ifthenelse{\boolean{CLEANTITLE}}{%
	\pgfplotsset{every axis post/.append style={title = {} }}%
}{}%
\provideboolean{CLEARYLABEL}\ifthenelse{\boolean{CLEARYLABEL}}{%
	\pgfplotsset{every axis post/.append style={ylabel = {} }}%
}{}%
\provideboolean{NOLEGEND}%
\providecommand{\thisYlabelopacity}{1.0}%
\provideboolean{CLEANXAXIS}\ifthenelse{\boolean{CLEANXAXIS}}{%
	\pgfplotsset{every axis post/.append style={xticklabels = {} }}%
}{}%
\provideboolean{CLEANYAXIS}\ifthenelse{\boolean{CLEANYAXIS}}{%
	\pgfplotsset{every axis post/.append style={ylabel = {} }}%
}{}%
\providecommand{\figheight}{310pt}%
\provideboolean{CLEANXAXIS}\ifthenelse{\boolean{CLEANXAXIS}}{%
	\pgfplotsset{every axis post/.append style={xlabel = {} }}%
}{}%

\definecolor{darkcyan27158119}{RGB}{27,158,119}
\definecolor{darkgray176}{RGB}{176,176,176}

\begin{axis}[
clip=false,
every axis x label/.append style={opacity=\thisXlabelopacity},
every axis y label/.append style={opacity=\thisYlabelopacity},
height=\figheight,
tick align=outside,
tick pos=left,
width=\figwidth,
x grid style={darkgray176},
xlabel={spikes/bin},
xmajorgrids,
xmin=-0.49, xmax=1.49,
xtick style={color=black},
xtick={0.0,1.0},
y grid style={darkgray176},
ylabel={\(\displaystyle \log_2\)(count)},
ymajorgrids,
ymin=0, ymax=25.175540805524,
ytick style={color=black}
]
\draw[draw=none,fill=darkcyan27158119,fill opacity=0.9] (axis cs:-0.4,0) rectangle (axis cs:0.4,23.9767055290705);
\draw[draw=none,fill=darkcyan27158119,fill opacity=0.9] (axis cs:0.6,0) rectangle (axis cs:1.4,18.3730605478216);
\ifthenelse{\boolean{NOLEGEND}}{\legend{}}{}
\end{axis}

\end{tikzpicture}
                    \end{minipage}
                \end{tabular}
            }
        \end{minipage}
        \hfill
        \pgfmathparse{(\pngwidth + \dmfcwidth)/\denominator}%
        \begin{minipage}[c]{\pgfmathresult\FullTextWidth}\vspace{0in}
            \subfloat[][]{%
                \label{subfig:dmfc}%
                \begin{tabular}[b]{c}
                    \dmfcrsg: Dorsomedial frontal cortex\\
                    \pgfmathparse{\pngwidth/\denominator}
                    \begin{minipage}[t]{\pgfmathresult\FullTextWidth}\vspace{0in}
                        \includegraphics[width=\linewidth]{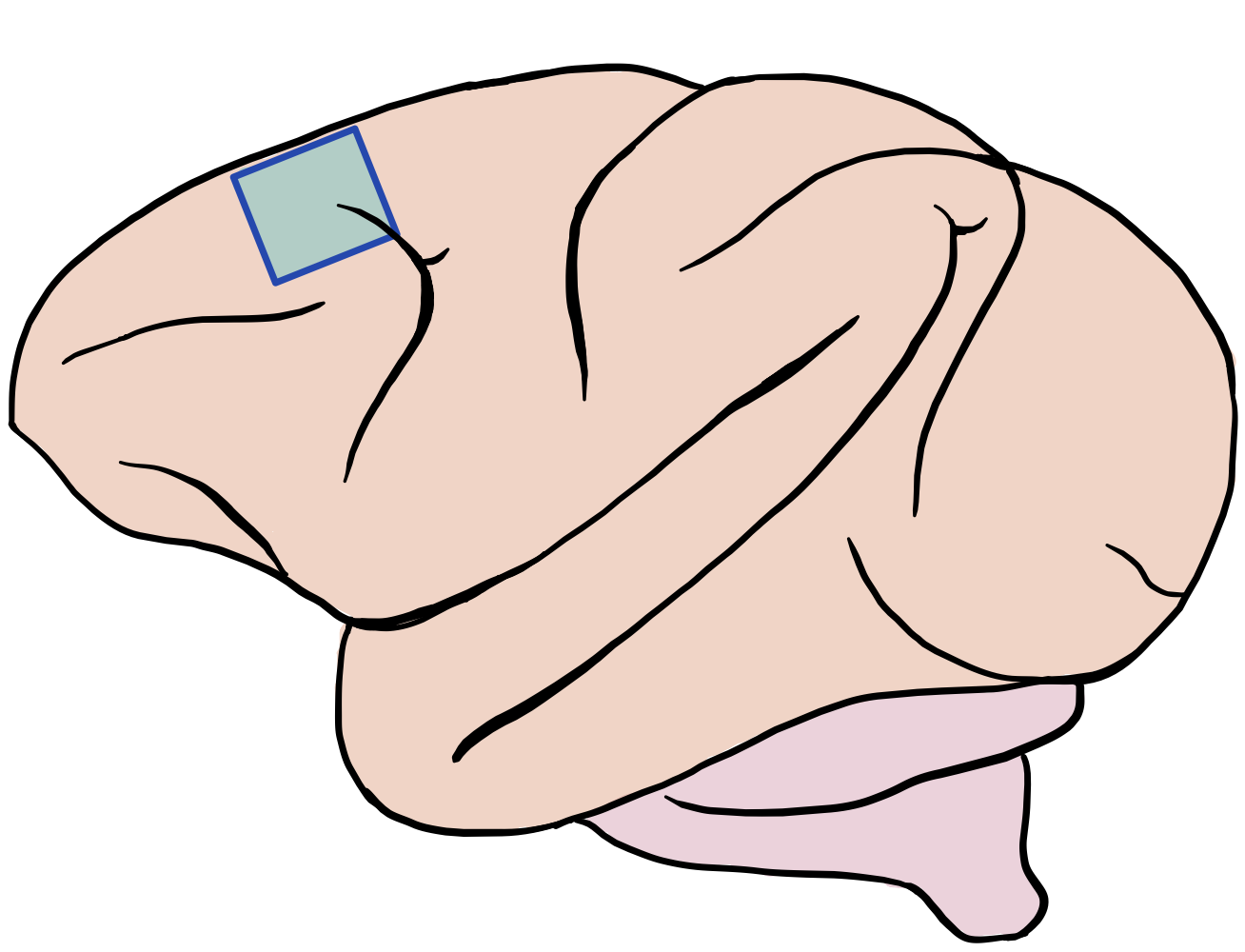}\\
                        \includegraphics[width=\linewidth]{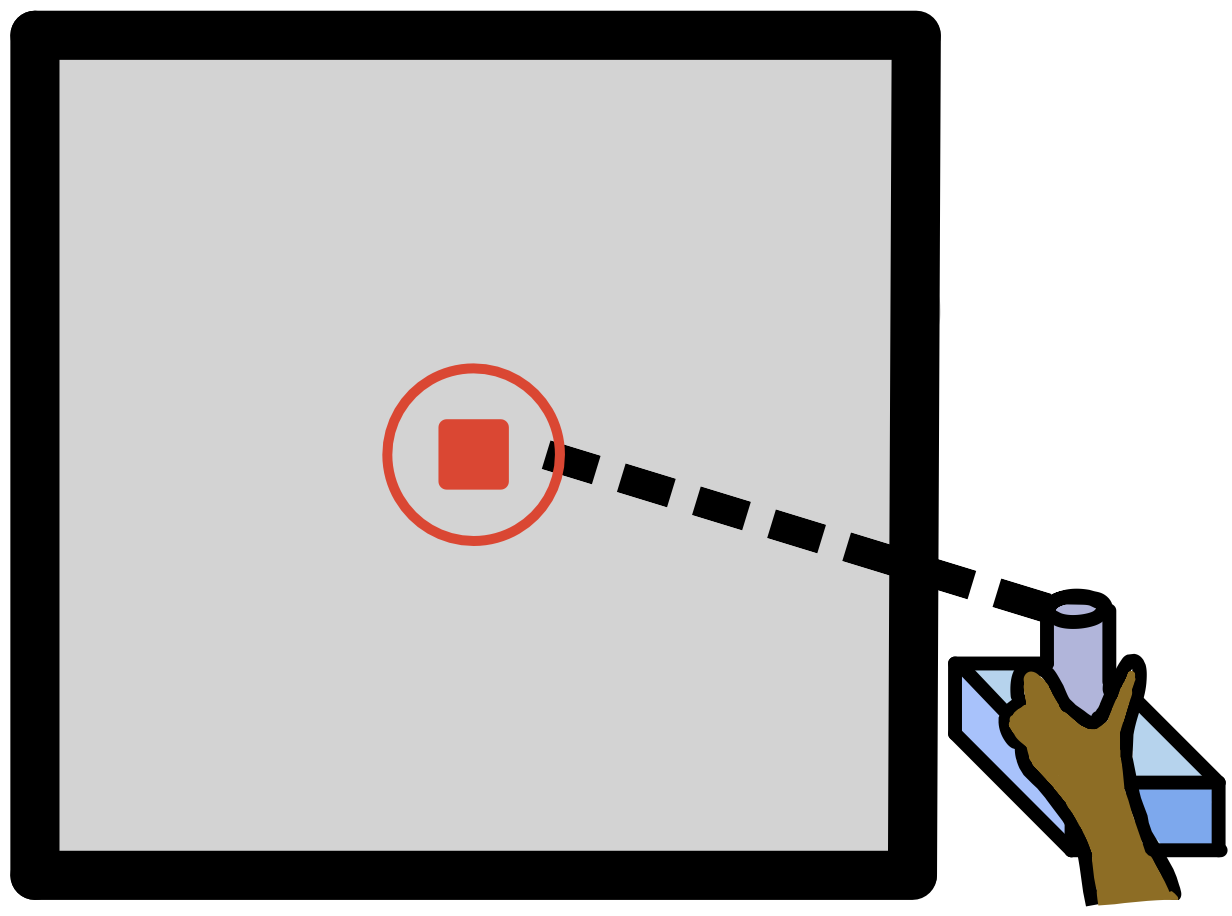}    
                    \end{minipage}
                    \pgfmathparse{\dmfcwidth/\denominator}
                    \begin{minipage}[t]{\pgfmathresult\FullTextWidth}\vspace{0in}
                        \scriptsize%
                        \providecommand{\figheight}{2in}%
                        \providecommand{\figwidth}{\linewidth}%
\begin{tikzpicture}
\providecommand{\thisXlabelopacity}{1.0}%
\providecommand{\figwidth}{360pt}%
\pgfplotsset{compat=1.15}%
\provideboolean{CLEARXLABEL}\ifthenelse{\boolean{CLEARXLABEL}}{%
	\pgfplotsset{every axis post/.append style={xlabel = {} }}%
}{}%
\provideboolean{CLEANYAXIS}\ifthenelse{\boolean{CLEANYAXIS}}{%
	\pgfplotsset{every axis post/.append style={yticklabels = {} }}%
}{}%
\provideboolean{CLEANTITLE}\ifthenelse{\boolean{CLEANTITLE}}{%
	\pgfplotsset{every axis post/.append style={title = {} }}%
}{}%
\provideboolean{CLEARYLABEL}\ifthenelse{\boolean{CLEARYLABEL}}{%
	\pgfplotsset{every axis post/.append style={ylabel = {} }}%
}{}%
\provideboolean{NOLEGEND}%
\providecommand{\thisYlabelopacity}{1.0}%
\provideboolean{CLEANXAXIS}\ifthenelse{\boolean{CLEANXAXIS}}{%
	\pgfplotsset{every axis post/.append style={xticklabels = {} }}%
}{}%
\provideboolean{CLEANYAXIS}\ifthenelse{\boolean{CLEANYAXIS}}{%
	\pgfplotsset{every axis post/.append style={ylabel = {} }}%
}{}%
\providecommand{\figheight}{310pt}%
\provideboolean{CLEANXAXIS}\ifthenelse{\boolean{CLEANXAXIS}}{%
	\pgfplotsset{every axis post/.append style={xlabel = {} }}%
}{}%

\definecolor{darkgray176}{RGB}{176,176,176}
\definecolor{deeppink23141138}{RGB}{231,41,138}

\begin{axis}[
clip=false,
every axis x label/.append style={opacity=\thisXlabelopacity},
every axis y label/.append style={opacity=\thisYlabelopacity},
height=\figheight,
tick align=outside,
tick pos=left,
width=\figwidth,
x grid style={darkgray176},
xlabel={spikes/bin},
xmajorgrids,
xmin=-0.64, xmax=4.64,
xtick style={color=black},
xtick={0.0,1.0,2.0,3.0,4.0},
y grid style={darkgray176},
ylabel={\(\displaystyle \log_2\)(count)},
ymajorgrids,
ymin=0, ymax=24.8365513193379,
ytick style={color=black}
]
\draw[draw=none,fill=deeppink23141138,fill opacity=0.9] (axis cs:-0.4,0) rectangle (axis cs:0.4,23.6538583993694);
\draw[draw=none,fill=deeppink23141138,fill opacity=0.9] (axis cs:0.6,0) rectangle (axis cs:1.4,18.9975295910906);
\draw[draw=none,fill=deeppink23141138,fill opacity=0.9] (axis cs:1.6,0) rectangle (axis cs:2.4,13.4514687129474);
\draw[draw=none,fill=deeppink23141138,fill opacity=0.9] (axis cs:2.6,0) rectangle (axis cs:3.4,8.21431912080077);
\draw[draw=none,fill=deeppink23141138,fill opacity=0.9] (axis cs:3.6,0) rectangle (axis cs:4.4,1);
\ifthenelse{\boolean{NOLEGEND}}{\legend{}}{}
\end{axis}

\end{tikzpicture}
                    \end{minipage}
                \end{tabular}
            }
        \end{minipage}
        \begin{minipage}[t]{0.48\linewidth}\vspace{0in}
            \flushleft%
            \subfloat[][]{%
                \label{subfig:all_datasets_average_counts}%
                \centering\scriptsize%
                \providecommand{\figheight}{2.0in}%
                \providecommand{\figwidth}{0.9\linewidth}%
\begin{tikzpicture}
\providecommand{\thisXlabelopacity}{1.0}%
\providecommand{\figwidth}{360pt}%
\pgfplotsset{compat=1.15}%
\provideboolean{CLEARXLABEL}\ifthenelse{\boolean{CLEARXLABEL}}{%
	\pgfplotsset{every axis post/.append style={xlabel = {} }}%
}{}%
\provideboolean{CLEANYAXIS}\ifthenelse{\boolean{CLEANYAXIS}}{%
	\pgfplotsset{every axis post/.append style={yticklabels = {} }}%
}{}%
\provideboolean{CLEANTITLE}\ifthenelse{\boolean{CLEANTITLE}}{%
	\pgfplotsset{every axis post/.append style={title = {} }}%
}{}%
\provideboolean{CLEARYLABEL}\ifthenelse{\boolean{CLEARYLABEL}}{%
	\pgfplotsset{every axis post/.append style={ylabel = {} }}%
}{}%
\provideboolean{NOLEGEND}%
\providecommand{\thisYlabelopacity}{1.0}%
\provideboolean{CLEANXAXIS}\ifthenelse{\boolean{CLEANXAXIS}}{%
	\pgfplotsset{every axis post/.append style={xticklabels = {} }}%
}{}%
\provideboolean{CLEANYAXIS}\ifthenelse{\boolean{CLEANYAXIS}}{%
	\pgfplotsset{every axis post/.append style={ylabel = {} }}%
}{}%
\providecommand{\figheight}{310pt}%
\provideboolean{CLEANXAXIS}\ifthenelse{\boolean{CLEANXAXIS}}{%
	\pgfplotsset{every axis post/.append style={xlabel = {} }}%
}{}%

\definecolor{darkgray176}{RGB}{176,176,176}

\begin{axis}[
every axis x label/.append style={opacity=\thisXlabelopacity},
every axis y label/.append style={opacity=\thisYlabelopacity},
height=\figheight,
scaled x ticks=false,
tick align=outside,
tick pos=left,
width=\figwidth,
x grid style={darkgray176},
xlabel={mean spikes/bin/neuron},
xmajorgrids,
xmin=0, xmax=0.043963623046875,
xtick style={color=black},
xticklabel style={/pgf/number format/fixed, /pgf/number format/precision=3},
y grid style={darkgray176},
ymajorgrids,
ymin=-0.59, ymax=3.59,
ytick style={color=black},
ytick={0,1,2,3},
yticklabels={\dmfcrsg,\areabump,\mcrtt,\mcmaze}
]
\draw[draw=none,fill=black,fill opacity=0.9] (axis cs:0,-0.4) rectangle (axis cs:0.039794921875,0.4);
\draw[draw=none,fill=black,fill opacity=0.9] (axis cs:0,0.6) rectangle (axis cs:0.0418701171875,1.4);
\draw[draw=none,fill=black,fill opacity=0.9] (axis cs:0,1.6) rectangle (axis cs:0.0201568603515625,2.4);
\draw[draw=none,fill=black,fill opacity=0.9] (axis cs:0,2.6) rectangle (axis cs:0.0183563232421875,3.4);
\ifthenelse{\boolean{NOLEGEND}}{\legend{}}{}
\end{axis}

\end{tikzpicture}
            }%
        \end{minipage}
        \begin{minipage}[t]{0.48\linewidth}\vspace{0in}
            \centering%
            \subfloat[][]{%
                \label{subfig:Dataset_trials_bins_neurons}%
                \centering\scriptsize%
                \providecommand{\figheight}{2.0in}%
                \providecommand{\figwidth}{0.9\linewidth}%
\begin{tikzpicture}
\provideboolean{CLEANXAXIS}\ifthenelse{\boolean{CLEANXAXIS}}{%
	\pgfplotsset{every axis post/.append style={xticklabels = {} }}%
}{}%
\provideboolean{CLEANYAXIS}\ifthenelse{\boolean{CLEANYAXIS}}{%
	\pgfplotsset{every axis post/.append style={ylabel = {} }}%
}{}%
\providecommand{\figheight}{2in}%
\provideboolean{CLEANTITLE}\ifthenelse{\boolean{CLEANTITLE}}{%
	\pgfplotsset{every axis post/.append style={title = {} }}%
}{}%
\providecommand{\thisXlabelopacity}{1.0}%
\providecommand{\thisYlabelopacity}{1.0}%
\providecommand{\figwidth}{2in}%
\pgfplotsset{compat=1.15}%
\provideboolean{NOLEGEND}%
\provideboolean{CLEANXAXIS}\ifthenelse{\boolean{CLEANXAXIS}}{%
	\pgfplotsset{every axis post/.append style={xlabel = {} }}%
}{}%
\provideboolean{CLEANYAXIS}\ifthenelse{\boolean{CLEANYAXIS}}{%
	\pgfplotsset{every axis post/.append style={yticklabels = {} }}%
}{}%

\definecolor{darkgray176}{RGB}{176,176,176}

\begin{axis}[
clip=false,
colorbar,
colorbar style={ytick={60,80,100,120,140,160,180},yticklabels={100,150,,,,,},ylabel={no. neurons}},
colorbar style={ytick={60,80,100,120,140,160,180},yticklabels={60,80,100,120,140,160,180 }},
colormap={mymap}{[1pt]
  rgb(0pt)=(0,0,1);
  rgb(1pt)=(1,0,0);
  rgb(2pt)=(0,1,0)
},
every axis x label/.append style={opacity=\thisXlabelopacity},
every axis y label/.append style={opacity=\thisYlabelopacity},
height=\figheight,
point meta max=182,
point meta min=54,
tick align=outside,
tick pos=left,
width=\figwidth,
x grid style={darkgray176},
xlabel={no. trials},
xmin=267.45, xmax=2391.55,
xtick style={color=black},
xtick={0,500,1000,1500,2000,2500},
xticklabels={0,1000,2000,3000,2000,2500},
y grid style={darkgray176},
ylabel={trial length (samples)},
ymin=151, ymax=349,
ytick style={color=black}
]
\addplot [
  colormap={mymap}{[1pt]
  rgb(0pt)=(0,0,1);
  rgb(1pt)=(1,0,0);
  rgb(2pt)=(0,1,0)
},
  only marks,
  scatter,
  scatter src=explicit
]
table [x=x, y=y, meta=colordata]{%
x  y  colordata
2295 180 182.0
1080 160 130.0
364 160 65.0
1006 340 54.0
};
\draw (axis cs:2295,200) node[
  scale=1.1,
  fill=white,
  draw=black,
  line width=0.4pt,
  inner sep=2.9pt,
  fill opacity=1.0,
  anchor=south east,
  text=black,
  rotate=0.0
]{\mcmaze};
\draw (axis cs:1200,170) node[
  scale=1.1,
  fill=white,
  draw=black,
  line width=0.4pt,
  inner sep=2.9pt,
  fill opacity=1.0,
  anchor=south west,
  text=black,
  rotate=0.0
]{\mcrtt};
\draw (axis cs:364,175) node[
  scale=1.1,
  fill=white,
  draw=black,
  line width=0.4pt,
  inner sep=2.9pt,
  fill opacity=1.0,
  anchor=south west,
  text=black,
  rotate=0.0
]{\areabump};
\draw (axis cs:1120,330) node[
  scale=1.1,
  fill=white,
  draw=black,
  line width=0.4pt,
  inner sep=2.9pt,
  fill opacity=1.0,
  anchor=north west,
  text=black,
  rotate=0.0
]{\dmfcrsg};
\ifthenelse{\boolean{NOLEGEND}}{\legend{}}{}
\end{axis}

\end{tikzpicture}
            }%
        \end{minipage}
        \captioning{Data sets.}{%
        For each of the four data sets \subcaprange{mc_maze}{dmfc} in the Neural Latents Benchmark \cite{Pei2021}, we show the brain area recorded from (upper left), a depiction of the task (lower left), and a histogram of the number of spikes per 5-ms bin; cf.\ \tbl{databreakdown}.
        \subcapref{all_datasets_average_counts} Average spike count (per 5-ms bin) per neuron for all four data sets.
        \subcapref{Dataset_trials_bins_neurons}
        Properties of each of the four data sets.
        Number of trials includes the ``remote'' data.
        }
        \label{fig:datasets}
    \end{figure}

\TableDataSets

\FigDynamicMaskingEffects

\TableLeaderboard

\TableHyperparameters

    \begin{figure}[!ht]
        \scriptsize%
        \newcommand{\figwidth}{0.31\linewidth}%
        \newcommand{\figheight}{1.5in}%
        \input{\tikzdir/mc_maze_PSTH}
        \captioning{Peri-stimulus time histograms for \mcmaze.} 
        {``PSTHs''---i.e., time-aligned averages---for three neurons (rows) and five random conditions (colors) were constructed for the firing-rate predictions from all three architectures (columns).
        Shading indicates $\pm$ the standard error.
        For comparison, the true PSTH (constructed according to the procedure of \cite{Pei2021}) is shown in the first column.
        Note that this procedure smooths the spike counts, so the true PSTH will typically be smoother than the model PSTHs (and standard errors are incomparable, so they have been omitted).
        All PSTHs were constructed on \emph{held-in neurons and time steps}, but from \emph{held-out (evaluation) trials}.
        }
        \label{fig:mc_maze_PSTHs}
    \end{figure}

    \begin{figure}[!ht]
        \scriptsize%
        \newcommand{\figwidth}{0.31\linewidth}%
        \newcommand{\figheight}{1.5in}%
        \input{\tikzdir/area2_bump_PSTH}
        \captioning{Peri-stimulus time histograms for \areabump.} 
        {``PSTHs''---i.e., time-aligned averages---for three neurons (rows) and five random conditions (colors) were constructed for the firing-rate predictions from all three architectures (columns).
        Shading indicates $\pm$ the standard error.
        For comparison, the true PSTH (constructed according to the procedure of \cite{Pei2021}) is shown in the first column.
        Note that this procedure smooths the spike counts, so the true PSTH will typically be smoother than the model PSTHs (and standard errors are incomparable, so they have been omitted).
        All PSTHs were constructed on \emph{held-in neurons and time steps}, but from \emph{held-out (evaluation) trials}.
        }
        \label{fig:area2_bump_PSTHs}
    \end{figure}

    \begin{figure}[!ht]
        \scriptsize%
        \newcommand{\figwidth}{0.31\linewidth}%
        \newcommand{\figheight}{1.5in}%
        \input{\tikzdir/dmfc_rsg_PSTH}
        \captioning{Peri-stimulus time histograms for \dmfcrsg.} 
        {``PSTHs''---i.e., time-aligned averages---for three neurons (rows) and five random conditions (colors) were constructed for the firing-rate predictions from all three architectures (columns).
        Shading indicates $\pm$ the standard error.
        For comparison, the true PSTH (constructed according to the procedure of \cite{Pei2021}) is shown in the first column.
        Note that this procedure smooths the spike counts, so the true PSTH will typically be smoother than the model PSTHs (and standard errors are incomparable, so they have been omitted).
        All PSTHs were constructed on \emph{held-in neurons and time steps}, but from \emph{held-out (evaluation) trials}.
        }
        \label{fig:dmfc_rsg_PSTHs}
    \end{figure}

\FigOptuna{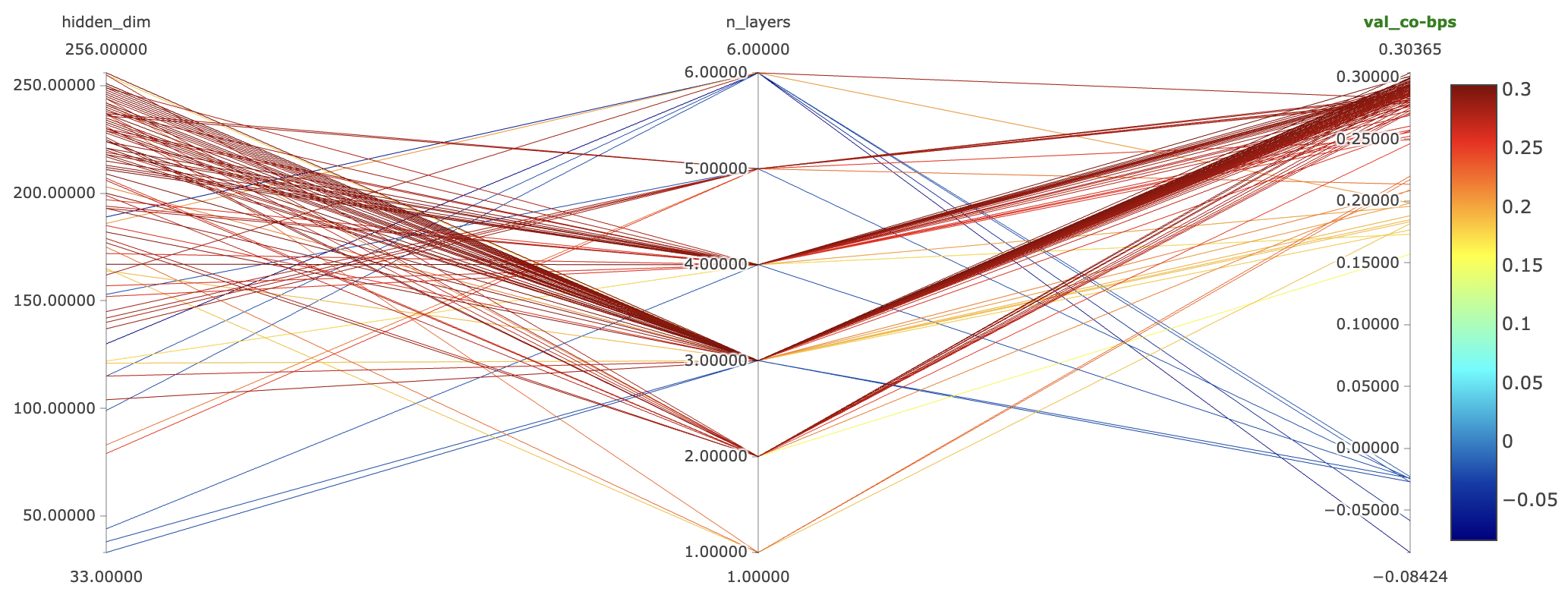}{\RNN}

\FigOptuna{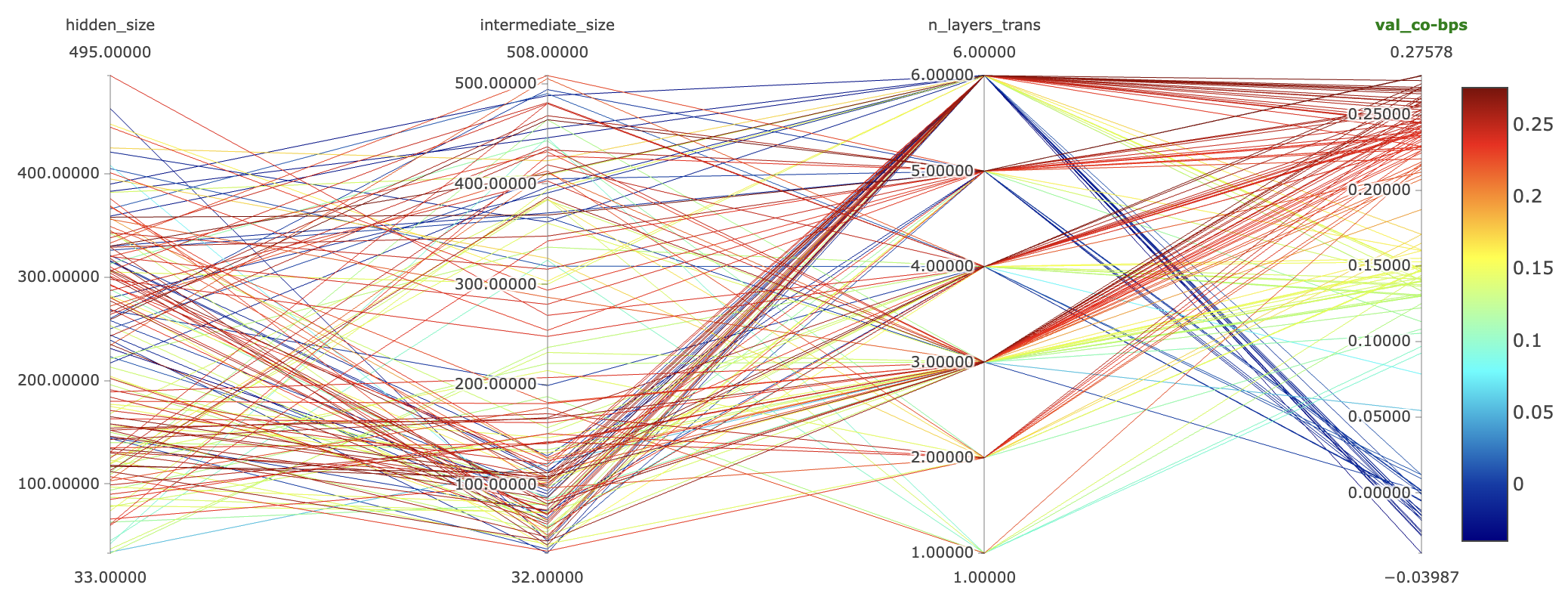}{\transformer}

\FigOptuna{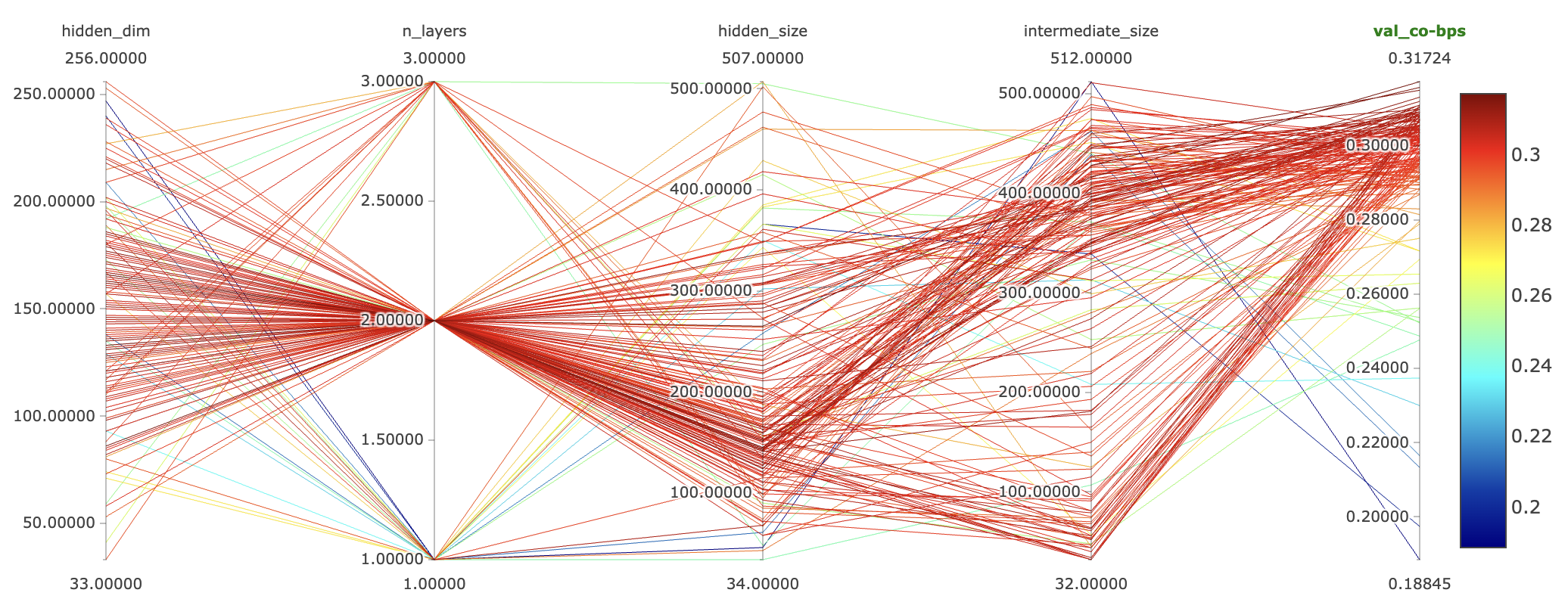}{\TERN}

\end{document}